\documentclass[twocolumn,           
               showpacs,            
               preprintnumbers,     
               aps,                 
               prd,          	    
               a4paper,             
               groupedaddress,      
               nofootinbib,         
               tightenlines,        
               floats,floatfix,     
               semlayer%
               ]{revtex4}

\usepackage{graphicx}
\usepackage{color}
\usepackage{latexsym}
\usepackage{amsmath,amssymb}        
\usepackage[draft=false]{hyperref}
\definecolor{myred}{rgb}{0.74,0.22,0.15}

\begin{document}



\title{Evolution of the Schr\"odinger--Newton system for a self--gravitating scalar field}

\author{F. Siddhartha Guzm\'an}
 \email{guzman@cct.lsu.edu}
\affiliation{Center for Computation and Technology, Louisiana State University. Baton Rouge, LA 70803\footnote{Current address}\\
and Max Planck Institut f\"ur Gravitationsphysik, Albert Einstein Institut, Am M\"uhlenberg 1, 14476 Golm, Germany.}
\author{L. Arturo Ure\~na--L\'opez}%
 \email{lurena@fisica.ugto.mx}
\affiliation{Instituto de F\'isica de la Universidad de Guanajuato, A.P. 150, C.P. 37150, Le\'on, Guanajuato, M\'exico.}%
\date{\today}
\pacs{04.40.-b, 98.35.Jk, 98.62.Gq}


\begin{abstract}
Using numerical techniques, we study the collapse of a scalar field configuration in the Newtonian limit of the spherically symmetric Einstein--Klein--Gordon (EKG) system, which results in the so called Schr\"odinger--Newton (SN) set of equations. We present the numerical code developed to evolve the SN system and topics related, like equilibrium configurations and boundary conditions. Also, we analyze the evolution of different initial configurations and the physical quantities associated to them. In particular, we readdress the issue of the gravitational cooling mechanism for Newtonian systems and find that all systems settle down onto a $0$--node equilibrium configuration.
\end{abstract}

\maketitle


\section{Introduction}
\label{sec:introduction}
In a previous paper of ours\cite{fsglau}, we studied the formation of a gravitationally bounded object comprised of scalar particles, under the influence of Newtonian gravity. The dynamics of the system is described by the coupled Schr\"odinger--Newton (SN) system of equations, which is nothing but the weak field limit of its general relativistic counterpart, the Einstein--Klein--Gordon (EKG) system. 

As at the moment we have no hints to finding an analytic solution for the evolution, we found necessary to develop numerical techniques to study the formation process of the scalar objects. The study of the dynamical properties of the fully time-dependent SN system has been done before in the literature\cite{seidel90,seidel94,schmethod,hu}, but more is needed in order to understand the gravitational collapse of a weakly gravitating scalar field.

The main aim of this paper is to perform an exhaustive numerical study of the collapse and evolution of a spherically symmetric scalar object in the Newtonian regime. Here, we develop a numerical strategy to evolve the SN system, and study important issues like the stability and the formation process of gravitationally bound scalar systems, a topic that has recently become attractive in Cosmology \cite{fsglau,schmethod,hu,phi2,dmota,jcerv,fuchs}.

A summary of the paper is as follows. In Sec.~\ref{mathematics}, we present the relativistic EKG and Newtonian SN equations that describe the evolution of a self-gravitating scalar field in the spherically symmetric case. Correspondingly, it is described how the EKG and the SN equations are solved in order to obtain regular and asymptotically flat solutions. These solutions are known as \textit{boson stars} and \textit{oscillatons}, respectively, for complex and real scalar fields. However, we focus our attention on their corresponding weak field limit, SN system and its properties.

In Sec.~\ref{sec:numerics}, we present an appropriate numerical code to evolve the SN system, providing details about the algorithms used. The issues concerning the physical boundary conditions imposed on the SN system and the accuracy of the code are of particular importance.

The results of the numerical evolutions are given in Sec.~\ref{results}. We systematically test the boundary conditions, the convergence of the numerical solutions and how the latter reproduces the expected results for the equilibrium configurations of the SN system.

Sec.~\ref{sec:cooling} is devoted to the study of the gravitational cooling mechanism, first described in\cite{seidel94}. Finally, the conclusions are collected in Sec.~\ref{conclusions}.

\section{Mathematical background}
\label{mathematics}
Here we describe the mathematical basis to deal with classical scalar fields, complex and real, in both the relativistic and Newtonian limits. The latter is given in much more detail as it will be our system of interest for the rest of this paper. 

To begin with, we write the equations that describe a scalar system within General Relativity, which are the coupled Einstein--Klein--Gordon equations (EKG). For simplicity, we deal only with the spherically symmetric case for a single scalar fluctuation. Then, we describe how the EKG equations can be solved to obtain regular and asymptotically flat solutions, which are known in the literature as boson stars and oscillatons. Some comments on the stability of these relativistic solutions are also given.

After that, we obtain the Newtonian limit of the EKG system through a post--Newtonian procedure, which yields the so called Schr\"odinger--Newton (SN) equations. The Newtonian limit is quite interesting by itself because many physical quantities can be defined and measured, quantities that are useful to describe the system in a detailed manner. Moreover, the SN system obeys a scale transformation such that the study of all the possible equilibrium configurations is reduced to the study of properly sized profiles, and this includes the evolution process too.

The SN system can be solved to find stationary solutions that we shall call Newtonian equilibrium configurations. These are called \textit{Newtonian boson stars} and \textit{Newtonian oscillatons}, following the relativistic nomenclature. Though Newtonian oscillatons are described by a larger set of equations than Newtonian bosons stars, their dynamical evolution is governed by the same SN system.
Finally, we will also calculate the perturbations of the $0$--node equilibrium configurations. These perturbations are a particular imprint of $0$--node solutions of the SN system, and also indicate that the latter are intrinsically stable. This information will be useful for the numerical studies performed in the following sections.

\subsection{The Einstein--Klein--Gordon equations}
The energy--momentum tensor of a complex scalar field $\Phi^{\textrm{(c)}}$ endowed with a scalar potential $V(|\Phi^\textrm{(c)}|)=m^2 |\Phi^\textrm{(c)}|^2$, reads
\begin{eqnarray}
T^{\rm (c)}_{\mu \nu} &=& \frac{1}{2} \left[ \Phi^\textrm{(c)}_{,\mu} \Phi^{\textrm{(c)} \ast}_{,\nu} + \Phi^{\textrm{(c)} \ast}_{,\mu} \Phi^\textrm{(c)}_{,\nu} \right. \nonumber \\
&& \left. - g_{\mu \nu} \left( \Phi^{\textrm{(c)} ,\sigma} \Phi^{\textrm{(c)}\ast}_{,\sigma} + m^2 |\Phi^\textrm{(c)}|^2 \right) \right] \, ; \label{set_complex}
\end{eqnarray}
whereas for a real scalar field $\Phi^\textrm{(r)}$ endowed with a potential $V(\Phi^\textrm{(r)})=(1/2)m^2 \Phi^{\textrm{(r)}2}$, is given by
\begin{equation}
T^{\rm (r)}_{\mu \nu} =  \Phi^\textrm{(r)}_{,\mu} \Phi^\textrm{(r)}_{,\nu} - \frac{1}{2}g_{\mu \nu} \left( \Phi^{\textrm{(r)},\sigma} \Phi^\textrm{(r)}_{,\sigma} + m^2 \Phi^{\textrm{(r)}2} \right) \, . \label{set_real}
\end{equation}
The parameter $m$ is interpreted in both cases as the mass of the scalar particles.

As we will work on the Newtonian limit of the scalar fields, we represent 
them using the same variables since much of the analytical treatment is similar 
for both type of fields. It is, however, easy to distinguish each case, as we 
will refer to complex fields wherever a complex conjugation appears in the formulas.

The equation of motion governing the evolution of the scalar field is the Klein-Gordon (KG) equation, which appears from the conservation of the energy--momentum tensor ${T^{{\rm (c,r)} \mu \nu}}_{;\nu}=0$. Hence, the KG equation is written, respectively for the complex and real cases, as
\begin{equation}
\Box \Phi^{\rm (c,r)} - m^2 \Phi^{\rm (c,r)} = 0 \, ,\label{kg} 
\end{equation}

\noindent where $\Box = \frac{1}{\sqrt{-g}}\partial_{\mu}[\sqrt{-g} g^{\mu \nu}\partial_{\nu} ]$ is the covariant d'Alambertian operator.

Within General Relativity, the evolution of the system is governed by the KG equation coupled to the Einstein equations. For simplicity, we consider the spherically symmetric case in the polar-areal slicing, for which the metric is written in the form
\begin{equation}
ds^2 = -\alpha^2 dt^2 + a^2 dr^2 + r^2 \left( d\theta^2 + \sin^2 \theta d\varphi^2 \right) \, ,
\label{metric}
\end{equation}

\noindent where $\alpha(t,r)$ and $a(t,r)$ are functions to be determined self-consistently from the matter distribution. The Einstein equations are written as usually, $G_{\mu \nu} = 8 \pi G T_{\mu \nu}$, being $G_{\mu \nu}$ the Einstein tensor corresponding to the metric Eq. (\ref{metric}). The term on the right hand side is the energy--momentum tensor~(\ref{set_complex}) or~(\ref{set_real}). Notice that we are using units in which $c=\hbar=1$, and then the Planck mass is given by $m_\textrm{Pl}=G^{-1/2}$.

\subsection{Relativistic equilibrium configurations}
Non-trivial solutions of the (spherically symmetric) EKG system exist that are regular everywhere, asymptotically flat, and for which the scalar field is confined to a finite region. These solutions are equilibrium configurations known as \textit{boson stars} and \textit{oscillatons}, whether we speak of complex or real scalar fields, respectively. We give here a brief description of both solutions.

\subsubsection{Boson stars}
Boson stars are the simplest solutions since the Klein--Gordon field admits an stationary solution of the form $\sqrt{4\pi G} \, \Phi^{\rm (c)}(t,r)= e^{-i\omega t}\phi(r)$, where $\phi(r)$ is a real function and $\omega$ is called the fundamental frequency of the boson star. Then, the metric functions are time--independent, which can be seen from the cancellation of the time dependence in the expression for the energy-momentum tensor (\ref{set_complex}). 

The EKG system becomes a set of coupled ordinary differential equations that has to be solved numerically. By imposing the conditions of regularity at the center, $a(r=0)=1$ and $\phi^\prime (r=0)=0$, together with the condition of asymptotic flatness, the EKG system becomes an eigenvalue problem. For each value of the field at the center, $\phi(r=0) \equiv \phi_0$, there are unique (eigen)values of $\omega$ and $\alpha(r=0)\equiv \alpha_0$ for which the conditions mentioned above are satisfied.

The resulting boson stars have been widely studied in the literature \cite{ruffini,colpi,liddle,pang,lee,moroz,gleiser,seidel90,seidel94,mielke,diegos,hawley1,hawley}, including the cases of non-spherically symmetric equilibrium configurations \cite{silveira,yoshida,ryan,kobayashi}. Their simple properties have also inspired some models of dark matter in galaxies. We will not comment more on boson stars, but the interested reader will find useful information in the references given above and in the reviews \cite{jetzer,mielkes}. 

\subsubsection{Oscillatons}
For the case of a real scalar field, it is not possible to propose regular and stationary solutions such that the metric functions are time--independent. Rather, it has to be taken into account that all fields are time--dependent. For this, it is usually assumed that a Fourier expansion of the field and the metric coefficients suffices to set up the behavior of the equilibrium configurations\cite{seidel91,luis,phi2}, as described next. 

First of all, we take the following Fourier expansions\cite{luis,phi2}
\begin{subequations}
\label{expansions}
\begin{eqnarray}
\sqrt{8\pi G} \, \Phi^{\rm (r)}(t,r) &=& \sum^{\infty}_{j=1} \phi_{j}(r) \cos(j\omega t) \, , \label{expansionsa} \\
A(t,r) &=& \sum^{\infty}_{j=0} A_j (r) \cos(j\omega t) \,,\label{expansionsb} \\
C(t,r) &=& \sum^{\infty}_{j=0} C_j(r) \cos(j\omega t) \, , \label{expansionsc}
\end{eqnarray}
\end{subequations}
and substitute them into the EKG equations. Here, $A(t,r)=a^2(t,r)$, $C(t,r)=a^2 (t,r) / \alpha^2(t,r)$, and $\omega$ is the fundamental frequency of the system.

A set of coupled ordinary differential equations is obtained by setting each collected Fourier coefficient to zero. The non-linearity of the Einstein equations shows up because the different modes are mixed, and then the equations for all the modes have to be solved simultaneously. In order to avoid an infinite set of equations a cut-off mode is introduced, and thus expansions~(\ref{expansions}) are taken up to a maximum value $j=j_{\rm max}$, and then modes for which $j>j_{\textrm{max}}$ are eliminated by hand.

As for the boson star case, we impose the condition of regularity at the center, which now reads $\phi^\prime_j (r=0)=0$ and $a_0(r=0)=1 \, , a_{j \geq 1}(r=0)=0$. Again, we assume asymptotic flatness which converts the solution of the EKG system into an eigenvalue problem. The different solutions can be labeled by the value at the center of the first scalar field coefficient $\phi_1(r=0)\equiv \phi_0$, for which there is a set of eigenvalues of $\omega$, $\phi_{j\geq 2}(r=0)\equiv \phi^{j}_0$, and $C_j(r=0)\equiv C^{j}_0$ found to satisfy the boundary conditions. In practice, only the odd modes of the scalar field and the even modes of the metric functions are non trivial.

As would have been guessed, the properties of oscillatons are similar to those of boson stars, and this has motivated the inclusion of oscillatons as models of galaxy dark matter \cite{gal}. However, they have not been as exhaustively studied as boson stars, and many of their properties may remain undiscovered. 

\subsubsection{Stability}
Stability of boson stars has been studied both numerically and analytically\cite{gleiser,seidel90}. Restricting ourselves to the ground state configurations (also called 0--node solutions), it has been shown that there are stable equilibrium configurations, generating the \textit{S-branch}, which are stable against radial perturbations. That they are stable is also indicated by the fact that S-configurations are indeed final states for a wide variety of initial configurations -including those in excited states- settle down onto at the end of numerical evolutions. Other initial configurations either collapse to form a black hole or disperse away to infinity.

On the other hand, oscillatons have been only barely studied and it is commonly believed that they are intrinsically unstable. Though there is not an analytical proof for the intrinsic stability of oscillatons, numerical evolutions have shown that oscillatons can be classified into unstable and stable configurations, the latter being called S-branch oscillatons, following the boson star nomenclature. S-oscillatons are stable against radial perturbations, and they indeed play the role of final states of evolved systems made of real scalar fields \cite{phi2}. However, this is not a rigorous proof of the stability of oscillatons, and it is still possible that they have a long life time and decay very slowly\cite{dpage}.

\subsection{The Newtonian limit}

Now, we are interested in the weak field limit version of the coupled Einstein-Klein-Gordon (EKG) equations, for which $|\sqrt{4\pi G} \Phi^{\rm (c,r)}(t,r)| \ll 1$ and, correspondingly, $|\alpha^2(t,r)-1| \ll 1$ and $|a^2(t,r)-1| \ll 1$. It turns out that the weak field limit of boson stars and oscillatons are quite similar. A common feature that simplifies the calculations is that weak equilibrium configurations have only one frequency, $\omega \simeq m$, whose corresponding Fourier mode is the dominant one in the case of oscillatons. 

As it was shown in~\cite{seidel90,fsglau}, the weak field version of the EKG system is found through the standard post--Newtonian treatment, for both complex and real scalar fields. For this reason, we shall call them Newtonian boson stars and Newtonian oscillatons, respectively.

\subsubsection{Newtonian boson stars} 

We start with the complex case following\cite{seidel90}. First, we express the scalar field and metric coefficients in terms of the Newtonian fields $\psi,U,A$ as\begin{subequations}
\label{newtbos}
\begin{eqnarray}
\Phi^{\rm (c)}(\tau,x) &=& \frac{1}{\sqrt{4\pi G}} e^{-i\tau}\psi(\tau,x) \, , \label{newtbos1} \\
\alpha^2 (\tau,x)&=& 1+2U(\tau,x) \, , \label{newtbos2} \\
a^2 (\tau,x) &=& 1+2A(\tau,x) \, , \label{newtbos3}
\end{eqnarray}
\end{subequations}
where we have also introduced the dimensionless quantities $\tau = mt, ~ x=mr$. Notice that $U(\tau,x)$ and $A(\tau,x)$ are real valued fields. 

Next, we assume that the new fields are of order ${\cal O}(\epsilon^2) \ll 1$, and that their derivatives obey the standard post-Newtonian rules $\partial_\tau \sim \epsilon \partial_x \sim {\cal O}(\epsilon^4)$. Therefore, considering the leading order terms only, the EKG equations become
\begin{eqnarray}
i\partial_\tau \psi &=& -\frac{1}{2x} \partial^2_x (x\psi) + U \psi \, ,  \label{schroedinger}\\
\partial^2_x (xU) &=& x \psi \psi^\ast \, , \label{poisson}\\
\partial_x (x A) &=& x^2 \psi \psi^\ast \, .\label{extra1}
\end{eqnarray}

Notice that $\psi(\tau,x)$ is a typical complex Schr\"odinger wave function, now coupled to $U(\tau,x)$, which is the usual Newtonian gravitational potential. For this reason, Eqs.~(\ref{schroedinger}) and~(\ref{poisson}) are named the Schr\"odinger--Newton (SN) system~\cite{lee,pang,moroz,seidel90,luis}.

\subsubsection{Newtonian oscillatons}
The post--Newtonian approximation for oscillatons proceeds in a similar manner as for the complex scalar field. However, to take into account the oscillating nature of oscillatons we need to take extra fields, and then we should consider the expansions
\begin{subequations}
\label{newtoscillaton}
\begin{eqnarray}
\Phi^{\rm (r)}(\tau,x) &=& \frac{1}{\sqrt{8\pi G}} \left[ e^{-i\tau} \psi(\tau,x) + {\rm c.c.} \right] \, , \label{phi_expansion}\\
\alpha^2 (\tau,x)&=& 1+2U(\tau,x) + e^{-2i\tau} U_2(\tau,x) + {\rm c.c.} \, , \label{gtt_expansion}\\
a^2 (\tau,x) &=& 1+2A(\tau,x) + e^{-2i\tau} A_2(\tau,x) + {\rm c.c.} \label{grr_expansion}
\end{eqnarray}
\end{subequations}
Notice that the fields $\psi(\tau,x)$, $U(\tau,x)$ and $A(\tau,x)$ have the same interpretation as in Eqs.~(\ref{newtbos}). But, this time we have introduced two new {\it complex} fields $U_2(\tau,x)$ and $A_2(\tau,x)$ to take into account the oscillating nature of oscillatons. 

In any case, all fields are considered to obey the post--Newtonian rules depicted above. Therefore, the evolution equations for $\psi(\tau,x)$, $U(\tau,x)$ and $A(\tau,x)$ are Eqs.~(\ref{schroedinger}),~(\ref{poisson}) and~(\ref{extra1}). In addition, an extra equation needed is
\begin{equation}
\partial_x U_2 = -x \psi^2 \, . \label{extra2}
\end{equation}

\subsubsection{Final remarks on the weak field limit}
Despite the presence of Eqs. (\ref{extra1}) and (\ref{extra2}), it should be noticed that the dynamical evolution of both Newtonian boson stars and Newtonian oscillatons is dictated by the SN system only, Eqs.~(\ref{schroedinger}) and~(\ref{poisson}). This only means that boson stars and oscillatons are quite similar at the weak field limit. 

So far, we have only used the $\{t,t\}$ and $\{r,r\}$ components of the Einstein equations to obtain Eqs.~(\ref{poisson}),~(\ref{extra1}) and~(\ref{extra2}). The $\{t,r\}$ component of the Einstein equations, in the weak field limit, reads
\begin{eqnarray}
\partial_\tau (\psi \psi^\ast) &=& \frac{i}{2x} \left[ \psi^\ast \partial^2_x(x\psi)- 
\psi \partial^2_x (x \psi^\ast) \right] \, , \label{current_conservation}\\
\partial_\tau A_2 &=& 2i A_2 - \frac{ix}{2} \psi \partial_x \psi \, .\label{extra3}
\end{eqnarray}

Eq.~(\ref{current_conservation}) is just the conservation of probability as we know it from Quantum Mechanics. Eq.~(\ref{extra3}) only appears for Newtonian oscillatons, and represents the oscillatory behavior of the metric coefficient $g_{rr}$. The rest of the Einstein components do not provide independent information.

We should also mention here what we mean by the \textit{weakness} of a Newtonian systems. The weakness of our configurations can be parametrized by the normalized value of the scalar field, and then we speak of the Newtonian limit if $|\sqrt{4\pi G} \Phi^{\rm (c,r)}(t,r)| \leq 10^{-3}$\cite{luis}. Any configuration with a larger value should be treated using the fully relativistic EKG system.

\subsection{Properties of the Schr\"odinger-Newton system}
\label{snproperties}
The properties of the scalar solitons, boson stars and oscillatons, can be more easily studied in the Newtonian limit, where we have a better understanding on the physical processes involved. 

To begin with, we can unambiguously define many physical quantities that will help us to study the evolution and formation of weak scalar solitons. Some of them are listed in Table~\ref{table1}.
\begin{table}
\caption{\label{table1} Quantities defined for the SN system.}
\begin{ruledtabular}
\begin{tabular}{lr}
Density of particles & $\rho(\tau,x) = \psi \psi^\ast \, ,$ \\
Mass number & $M(\tau,x) = \int^x_0 \rho y^2 \, dy \, ,$ \\
Kinetic energy & $K(\tau,x) = -(1/2) \int^x_0 \psi^\ast \partial^2_y (y \psi)  y \, dy \, ,$ \\
Gravitational energy & $W(\tau,x) = (1/2) \int^x_0 \rho U y^2 \, dy \, ,$ \\
Current of particles & $J(\tau,x) = (i/2) \left[ \psi \partial_x \psi^\ast - \psi^\ast \partial_x \psi \right] \, .$
\end{tabular}
\end{ruledtabular}
\end{table}
These quantities are directly linked to their relativistic counterparts, whenever the latter can be properly defined. For instance, the mass number $M(\tau ,x)$ is related to the total mass $M_{\rm T}$ of the soliton configurations through
\begin{equation}
M_{\rm T} = 4\pi \int^\infty_0 \rho_\Phi (\tau,x) x^2 dx \simeq \frac{m^2_{\rm Pl}}{m} M(\tau,x =\infty) \, , \label{massnumber}
\end{equation}
where $\rho_\Phi = - {T^t}_t$ in Eqs.~(\ref{set_complex}) and~(\ref{set_real}), for which we also find that $\rho_\Phi \simeq (1/4\pi) m^2 m^2_{\rm Pl} \psi \psi^\ast$ in the weak field limit.

\subsubsection{Scaling properties}
Finally, we want to point out that Eqs. (\ref{schroedinger}-\ref{poisson}) obey a scaling symmetry of the form\cite{fsglau}
\begin{equation}
\left\{ \tau,x,U,U_2,A,\psi \right\} \rightarrow \left\{ \lambda^{-2} \hat{\tau},\lambda^{-1} 
\hat{x}, \lambda^2 \hat{U}, \lambda^2 \hat{U}_2, \lambda^2 \hat{A}, \lambda^2 \hat{\psi} \right\} \, ,
\label{scale1}
\end{equation}
where $\lambda$ is an arbitrary parameter. 

This means that, once we have found a solution to the SN system, there is a complete family of solutions which are related one to each other just by a scaling transformation. Likewise, the quantities shown in Table~\ref{table1} obey the scaling transformation
\begin{equation}
\left\{ \rho , M, K, W, J \right\} \rightarrow \left\{ \lambda^{4} \hat{\rho},\lambda \hat{M}, 
\lambda^3 \hat{K}, \lambda^3 \hat{W}, \lambda^5 \hat{J} \right\} \, .
\label{scale2}
\end{equation}

Eq.~(\ref{scale1}) can be used to reduce significantly the space of possible equilibrium Newtonian configurations we need to study as follows. We will find solutions of the SN system for 'hat' quantities only, and then the 'non--hat' quantities will be obtained by applying the inverse of the scaling transformation~(\ref{scale1}). With this in mind, all quantities should be thought of as 'hat' quantities henceforth; we will write a 'hat' explicitly only when confusion can arise.

Furthermore, we can always solve the 'hat' system taking plainly that $|\hat{\psi}(0,0)|=1$. Thus, the weak field limit condition is translated into a constraint on the scale parameter as $\lambda^2 \leq 10^{-3}$.

Notice that the scaling transformation~(\ref{scale1}) does not apply for Eq.~(\ref{extra3}), but the latter indicates that $A_2(\tau,x)$ is, at least, $\lambda^2$--times smaller than the other Newtonian fields. In this respect, we can say that the $g_{rr}$ metric coefficient is time-independent for both kind of Newtonian configurations.

\subsection{Newtonian equilibrium configurations}
\label{subsec:neq}
The SN system can be solved to obtain non-singular self-gravitating configurations. These so called equilibrium configurations are of the form $\psi(\tau,x)=e^{-i\gamma \tau} \phi(x)$, for which Eqs.~(\ref{schroedinger}),~(\ref{poisson}),~(\ref{extra1}) and~(\ref{extra2}) become the following system of ordinary differential equations
\begin{subequations}
\label{unperturbed_ivp}
\begin{eqnarray}
(x \phi)_{,xx} &=& 2 x (U-\gamma) \phi \, , \label{unperturbed_ivpa} \\
(x U)_{,xx} &=& x \phi^2 \, , \label{unperturbed_ivpb} \\
(u_2)_{,x} &=& -x \phi^2 \, , \label{unperturbed_ivpc} \\
(x A)_{,x} &=& x^2 \phi^2 \, , \label{unperturbed_ivpd}
\end{eqnarray}
\end{subequations}
where $U_2(\tau,x) = e^{-2i\gamma \tau} u_2(x)$.

If we look for regular ($\phi_{,x} (0) =A(0)=0$) and finite configurations ($\phi (x\rightarrow \infty) \rightarrow 0$), the SN system becomes an eigenvalue problem. For $\phi(x=0)=1$, there are unique values of $\gamma$, $U(0)$ and $u_2(0)$, for which the boundary conditions are fulfilled. 

Although the system~(\ref{unperturbed_ivp}) has to be solved numerically, we can enunciate some analytical properties\cite{pang,moroz} that simplify the numerical treatment. Eqs.~(\ref{extra1}) and~(\ref{unperturbed_ivp}) can be formally integrated up to
\begin{subequations}
\label{integral_ivp}
\begin{eqnarray}
\phi (x) &=& 1 + \int^{x}_0 y(1-y/x) \left( u -\gamma \right) \phi \, dy \, \label{integral_ivpa} \\
U (x) &=& U (0) + \int^{x}_0 y \, \phi^2 dy -\frac{M(x)}{x} \, , \label{integral_ivpb} \\
u_2 (x) &=& u_2 (0) - \int^x_0 y \, \phi^2 dy \, , \label{integral_ivpc} \\
A(x) &=& \frac{M(x)}{x} \, . \label{integral_ivpd}
\end{eqnarray}
\end{subequations}
The relativistic condition of asymptotic flatness translates into $U(x\rightarrow \infty)=-M/x$ and $u_2(x\rightarrow \infty)=0$, from which we obtain
\begin{equation}
U (0)=-u_2 (0)= -\int^\infty_0 y \phi^2 dy \, . \label{outvals}
\end{equation}
 The numerical solutions are then found by using a shooting method to adjust the value of $\gamma$, which is the only free eigenvalue, since the other two can be considered {\it output} values through Eq.~(\ref{outvals})\cite{luis}. 

\begin{figure}[htp]
\includegraphics[width=8cm]{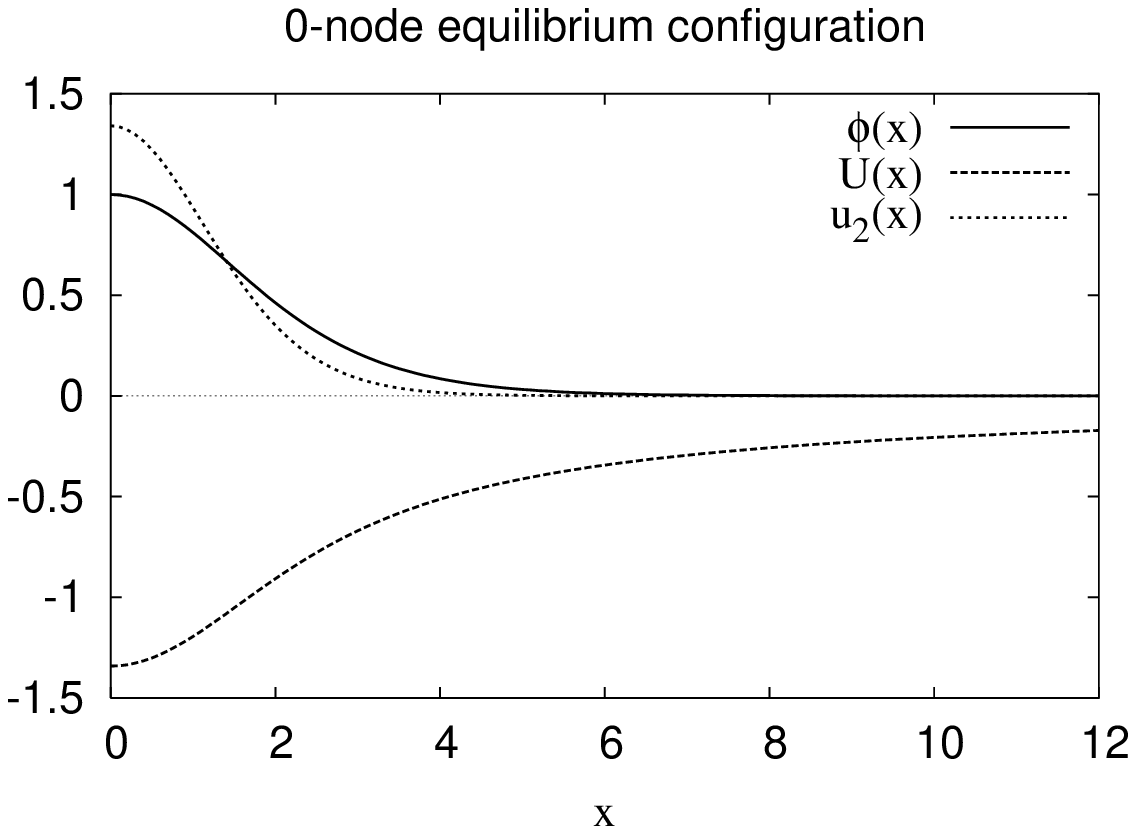}
\includegraphics[width=8cm]{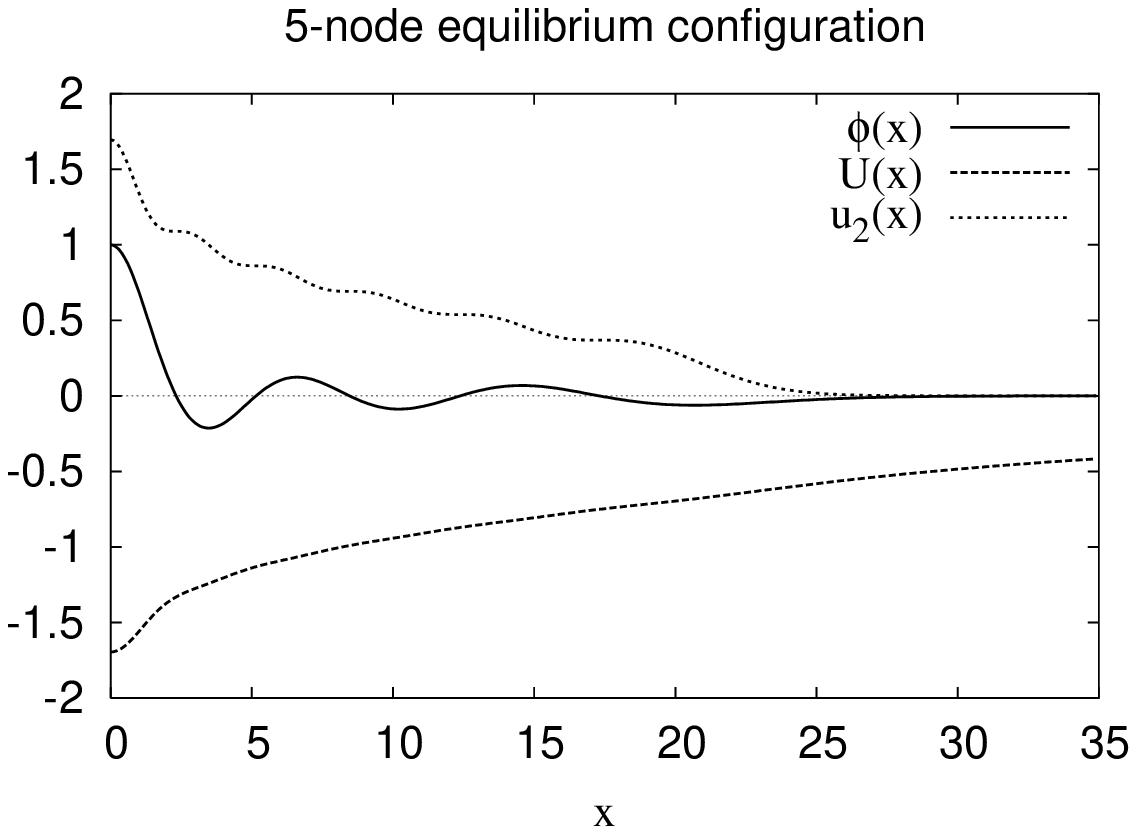}
\caption{\label{fig:f2} Profiles of $\phi(x)$, $U(x)$ and $u_2(x)$ for a 
$0$--node and a $5$--node Newtonian equilibrium configurations, see text and Table~\ref{table2} for details.} 
\end{figure}

In Fig.~\ref{fig:f2} we show the equilibrium configurations for a $0$--node and a $5$--node configurations, and the required eigenvalues for different node configurations are shown in Table~\ref{table2}.
\begin{table}
\caption{\label{table2} Eigenvalues of Newtonian equilibrium 
configurations. Shown also are their corresponding mass $M$, the $95$\% 
mass radius $x_{95}$, and the kinetic $K$ and gravitational $W$ energies. The precision of these numbers is in terms of the tolerance we used in our shooting routine to solve the initial value problem, with the boundaries placed at least three times $x_{95}$.}
\begin{ruledtabular}
\begin{tabular}{ccccccc}
Nodes & $\gamma$ & $U (x=0)$ & $M$ & $x_{95}$ & $K$ & $W$ \\
0 & $-0.69223$ & $-1.3418$ & $2.0622$ & $3.93 $ & $0.47585$ & $-0.95169$ \\
1 & $-0.64793$ & $-1.5035$ & $4.5874$ & $8.04 $ & $0.99037$ & $-1.9813$ \\
2 & $-0.63095$ & $-1.5811$ & $7.0969$ & $12.18$ & $1.4924$ & $-2.9850$ \\
3 & $-0.62081$ & $-1.6308$ & $9.5927$ & $16.35$ & $1.9850$ & $-3.9702$ \\
4 & $-0.61385$ & $-1.6670$ & $12.077$ & $20.54$ & $2.4706$ & $-4.9424$ \\
5 & $-0.60852$ & $-1.6954$ & $14.552$ & $24.74$ & $2.9520$ & $-5.9039$ \\
... & ...      & ...       & ...& ...   & ...   & ... \\
12 & $-0.58919$ & $-1.8060$ & $31.726$ & $54.40$ & $6.2333$ & $-12.463$ \\
... & ...      & ...       & ...& ...   & ...   & ...
\end{tabular}
\end{ruledtabular}
\end{table}

Some remarks are in turn. According to~(\ref{newtbos1}) and~(\ref{phi_expansion}), the fundamental angular frequency of the scalar field $\Phi^{\rm (c,r)}$ is given by $\omega /m = 1+\gamma$ with $\gamma = \lambda^2 \hat{\gamma}$. Due to the weakness restriction, $\lambda^2 \leq 10^{-3}$, $\Phi^{\rm (c,r)}$ has an angular frequency (in full units) of about $\omega \simeq m$, but such that $\omega /m \leq 1$, as it is the case for gravitationally bound configurations\cite{luis}.

From the Schr\"odinger equation~(\ref{schroedinger}), we find that the mass number ($M$), and the kinetic ($K$) and gravitational potential ($W$) energies are related through $\gamma M = K +2W$, while the total classical energy is $E_{\rm T}=K+W$. A common feature of any Newtonian equilibrium configuration is that, irrespectively of the number of nodes it may have, they all are virialized since $K/|W|=0.5$, see Table~\ref{table2}. Therefore, $E_{\rm T}=(1/2)W=(1/3)\gamma M < 0$, which also indicates that all of the equilibrium configurations are gravitationally bound objects.

\subsection{Analysis of perturbations}
\label{subsec:perturbations}
As we shall see later, our numerical simulations will be perturbed due to the discretization error, so we find illustrative to show how a first order perturbation model can predict extra modes that should appear in the evolution of equilibrium configurations. In this section, we restrict ourselves to the perturbations of $0$--node Newtonian equilibrium configurations.

An important concept here is that of \textit{quasinormal mode}. This is an oscillation mode which is characteristic of an equilibrium configuration which manifests when the system is perturbed.  In relativistic studies, once the quasinormal frequency is known the mass of the equilibrium configuration can be determined, thus it is possible to know the mass of the configuration a perturbed system is approaching to, without the need to follow the simulation until it settles down completely\cite{seidel90}. We shall show that the same can be done for weak field solitons.

First of all, we assume that
\begin{subequations}
\label{perturbed}
\begin{eqnarray}
\psi = \psi^{(0)} + \delta \psi && |\delta \psi| < 1 \label{perturbeda} \\
U   = U^{(0)} + \delta U && |\delta U| < 1 \label{perturbedb}
\end{eqnarray}
\end{subequations}
where $\psi^{(0)} = \phi(x)e^{-i\gamma \tau}$ and $U^{(0)}(x)$ are, respectively, the wave function and the gravitational potential calculated for the unperturbed system~(\ref{unperturbed_ivp}). Taking $\delta \psi = \varphi(x,t)e^{-i\gamma t}$, the equations for the perturbations read
\begin{subequations}
\label{perturbed_ev}
\begin{eqnarray}
i\partial_{\tau} \varphi &=& -\frac{1}{2x}\partial^2_{xx} (x \varphi) + \left( U^{(0)} - \gamma \right) \varphi + \delta U \phi \, , \label{perturbed_eva} \\
\partial^2_{xx} (x \delta U) &=& x \phi (\varphi^{*} + \varphi) \, , \label{perturbed_evb}
\end{eqnarray}
\end{subequations}
where we have neglected lower order terms. 

This system possesses similar scaling relations as those for the unperturbed SN: $\{\delta U, \varphi\} \rightarrow \{\lambda^4 \delta \hat{U}, \lambda^4 \hat{\varphi} \}$. Note that this indicates that the perturbations are suppressed faster than the unperturbed quantities as $\lambda \rightarrow 0$.

We then assume that the system~(\ref{perturbed_ev}) admits a stationary solution of the form $\varphi(x,t) = \varphi_1(x) e^{-i\sigma \tau} + \varphi_2(x) e^{i\sigma\tau}$, where $\varphi_1$ and $\varphi_2$ are real functions and $\sigma$ is a real frequency. To first order, the resulting perturbations of the wave function, the energy density and the number of particles read, respectively,
\begin{subequations}
\label{perturbations}
\begin{eqnarray}
\delta \psi(x,\tau) &=& e^{-i\gamma \tau} \left( \varphi_1 e^{-i\sigma\tau} + \varphi_2e^{i\sigma\tau} \right) \, , \label{perturbationsa} \\
\delta \rho (x,\tau) &=& \phi \left( \varphi_1 + \varphi_2 \right) \left( e^{-i\sigma \tau} + e^{i\sigma \tau} \right) \, , \label{perturbationsb} \\
\delta M (\tau,x) &=& \int^x_0 \delta \rho(y,\tau) \, y^2 dy \, , \label{perturbationsc}
\end{eqnarray}
\end{subequations}
and then we see that the perturbation of the gravitational potential admits the expansion $\delta U (\tau,x)= \delta u (x) (e^{-i\sigma\tau} + e^{i\sigma\tau})$. Hence, the perturbation equations~(\ref{perturbed_ev}) are further reduced to
\begin{subequations}
\label{perturbed_ivp}
\begin{eqnarray}
\partial^2_{xx} (x\varphi_1) &=& 2x \left( U^{(0)} - \gamma - \sigma \right) \varphi_1 + 2x\delta u \, \phi \, \label{perturbed_ivpa} \\
\partial^2_{xx} (x\varphi_2) &=& 2x \left( U^{(0)} - \gamma + \sigma \right)\varphi_2 + 2x\delta u \, \phi \, \label{perturbed_ivpb} \\
\partial^2_{xx} (x \delta u) &=& x \phi \left( \varphi_1 + \varphi_2 \right) \, .
\label{perturbed_ivpc}
\end{eqnarray}
\end{subequations}

The process followed to solve the system of equations~(\ref{perturbed_ivp}) is the same as that used to solve the unperturbed equations~(\ref{unperturbed_ivp}). As before, we will restrict ourselves again to 'hat' quantities only in an implicit manner, so that we fix the central value of one of the functions arbitrarily to $\varphi_1(0)=1$. The parameters $\varphi_2(0)$, $\delta u(0)$ and $\sigma$ are free. Then, a shooting method is used to fix the values of the free quantities provided the boundary conditions $\delta M (\tau,x\rightarrow \infty) = \varphi_1 (x\rightarrow \infty) = \varphi_2 (x\rightarrow \infty) = 0$, while at the same time we ask for regularity at the origin.

\begin{figure}[htp]
\includegraphics[width=8cm]{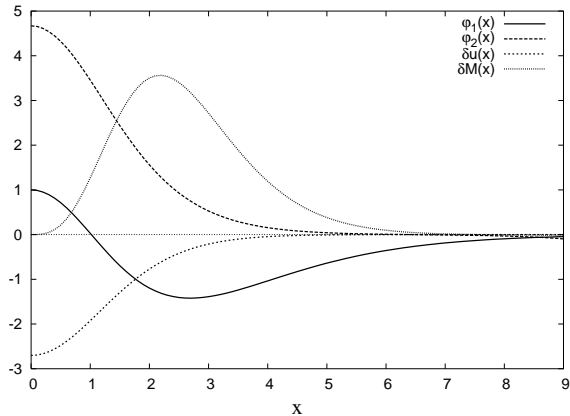}
\caption{\label{fig:perturbations} 
Profiles of linear perturbations of 
the wave function denoted by $\varphi_1$ and $\varphi_2$, and of the gravitational potential $\delta u$ for a 0--node equilibrium configuration, see Eqs.~(\ref{perturbed_ev}). These perturbation modes correspond to a quasinormal frequency $f=0.046$.} 
\end{figure}

The resulting perturbation profiles of $\varphi_1$, $\varphi_2$ and $\delta u$, are shown in Fig.~\ref{fig:perturbations}. The angular eigenfrequency of the perturbations is $\sigma=0.2916$, which coincides with other independent calculations\cite{hawley1,pang}\footnote{We should be careful when comparing values due to different scaling choices in the references. For\cite{pang}, take the first value of $\hat{\nu}_1$ in Table I, and then $\sigma=\hat{\nu}_1/\sqrt{2}$. As for\cite{hawley}, take the first value in Table II and use the conversion formula $\sigma=\sigma/\sqrt{6\times 10^{-2}}$.}.

The above solution means that the energy density and the gravitational potential of a 0--node configuration should oscillate with an angular frequency $\sigma$ when slightly perturbed (recall that the aforementioned quantities are strictly time--independent for an unperturbed configuration), and then $f=\sigma / (2\pi) \simeq 0.046$ would be the quasinormal frequency\footnote{This should be compared with the values given by the formula~(4.3) in~\cite{seidel90}, which in terms of 'hat' quantities reads
\begin{equation}
\hat{f} = \frac{\pi^2}{2\hat{x}^2_{95}}-\frac{\hat{M}}{\hat{x}_{95}} \, . \nonumber
\end{equation}
When applied to a 0--node equilibrium configurations, it gives the value $\hat{f}=0.033$, which is at variance with our previous calculations. Hereafter, we will refer to $\hat{f}=0.046$ as the correct value of the quasinormal mode of 0--node equilibrium configurations.} of the Newtonian boson system when perturbed. It should be noticed that the quasinormal frequency $f$ is absent in the wave function perturbation, but for this case, there are two perturbation frequencies, $\gamma \pm \sigma$.

The perturbation mode $\delta \psi$ in Eq.~(\ref{perturbeda}) is stationary since $\sigma$ is real. As it has been shown in\cite{pang}, this also means that the 0--node Newtonian configuration is stable against small radial perturbations. The stability analysis of excited equilibrium configurations will be performed with numerical simulations in the following sections.

\section{Numerical treatment}
\label{sec:numerics}
In this section we give explicit details of the implementation of a numerical code appropriate to solve the time--dependent SN system, Eqs.~(\ref{schroedinger}) and~(\ref{poisson}). The issues covered are the discretization of the system of differential equations, boundary conditions and the accuracy of the solutions.

\subsection{The evolution}
The time-dependent Schr\"odinger equation is solved using a fully implicit Crank-Nicholson scheme, which for equation~(\ref{schroedinger}) reads
\begin{equation}
\left(1+\frac{i}{2}H \Delta \tau \right) \psi^{n+1} = \left(1-\frac{i}{2} 
H \Delta \tau \right)\psi^{n}
\label{implicit}
\end{equation}

\noindent being $n$ the label of the time slice, $\Delta \tau$ is the separation between time slices and the Hamiltonian $H = -\frac{1}{2}\nabla^2 + U$ is written using second order centered finite differencing. This evolution scheme is found to be appropriate for the present problem since the approximation we use for the evolution operator is unitary, which allows one to preserve the number of particles $M=\int |\psi|^2 d^3 x$ \cite{nr}. 

The discretization of the Hamiltonian is second order accurate, and we handle the second term in the Laplacian as $\frac{2}{x}\frac{\partial \psi}{\partial x} = 4 \frac{\partial \psi}{\partial x^2}$, being the last one a derivative with respect to $x^2$ in order to avoid divergence at the origin. At the end, the finite differencing equation for the evolution of the wave function~(\ref{implicit}) is
\begin{eqnarray}
\left( \rho_1 - \frac{\rho_2}{x_j} \right) \psi^{n+1}_{j-1} + \left( -2\rho_1 - \frac{\Delta \tau U_j}{2} + i \right) \psi^{n+1}_{j} && \nonumber \\ +
\left( \rho_1 + \frac{\rho_2}{x_j} \right) \psi^{n+1}_{j+1} &=& \nonumber \\
\left( -\rho_1 + \frac{\rho_2}{x_j} \right) \psi^{n}_{j-1} + \left( 2\rho_1+\frac{\Delta \tau U_j}{2} + i \right) \psi^{n}_{j} && \nonumber \\
+ \left( -\rho_1 - \frac{\rho_2}{x_j} \right) \psi^{n}_{j+1} &&
\end{eqnarray}

\noindent where $\rho_1 = \Delta \tau/4 (\Delta x)^2$ and $\rho_2 = \Delta \tau/(4 \Delta x)$. The upper indices $n$ still label time slices and lower indices label spatial grid points with $j=0,1..N$, where $N$ labeling the last point of the grid. It is now defined a set of arrays that will help in the solution of this linear system at each time step,
\begin{subequations}
\begin{eqnarray}
a_j &=& \rho_1 - \frac{\rho_2}{x_j} \, , \\
b_j &=& -2\rho_1 - \frac{\Delta \tau U_j}{2} + i \, , \\
c_j &=& \rho_1 + \frac{\rho_2}{x_j} \, , \\
d_j &=& \left( -\rho_1 + \frac{\rho_2}{x_j} \right) \psi^{n}_{j-1} + \left( 2\rho_1+\frac{\Delta \tau U_j}{2}+i \right) \psi^{n}_{j} \nonumber \\
&& + \left( -\rho_1 - \frac{\rho_2}{x_j} \right) \psi^{n}_{j+1} \, ,
\end{eqnarray}
\end{subequations} 

\noindent for each grid point, with the special cases $a_0=0$, $c_N=0$, $d_0=\left( 2\rho_1+\frac{\Delta \tau U_j}{2} +i \right) \psi^{n}_{0}+\left( -\rho_1 - \frac{\rho_2}{x_j} \right) \psi^{n}_{1}$ and $d_N=\left( -\rho_1+\frac{\rho_2}{x_j} \right)\psi^{n}_{N-1} + \left( 2\rho_1 + \frac{\Delta \tau U_j}{2} + i \right) \psi^{n}_{N}$. Then, the following linear system of equations arises
\begin{eqnarray}
            b_0\psi^{n+1}_0 + c_0\psi^{n+1}_1 &=& d_0 \, , \nonumber\\
a_1\psi^{n+1}_0 + b^{n+1}_1\psi_1 + c_1\psi^{n+1}_2 &=& d_1 \, , \nonumber\\
&.& \, , \nonumber\\  
&.& \, , \nonumber\\
&.& \, , \nonumber\\
a_{N-1}\psi^{n+1}_{N-2} + b_{N-1}\psi^{n+1}_{N-1} + c_{N-1}\psi^{n+1}_N &=&
d_{N-1} \, , \nonumber\\
a_{N}\psi^{n+1}_{N-1} + b_{N}\psi^{n+1}_{N} &=& d_{N} \, . \nonumber
\end{eqnarray}
Such tridiagonal linear system is solved for the wave function $\psi^{n+1}$ at each time slice using a simple algorithm \cite{nr}.\\  

\subsection{The Poisson equation}
In order to solve the Poisson equation we write it down in the form $\partial^{2}_x (x U) = x |\psi|^2$ and integrate inwards from the outer boundary. For this we use the Numerov algorithm, whose finite differencing expression for the gravitational potential reads
\begin{eqnarray}
x^{n}_{j} U^{n}_{j} &=& 2x^{n}_{j+1}U^{n}_{j+1} - x^{n}_{j+1}U^{n}_{j+2} + \frac{(\Delta x)^2}{12} \times \nonumber \\
&& \left( x^{n}_{j+2}|\psi^{n}_{j+2}|^2 + 10 x^{n}_{j+1}|\psi^{n}_{j+1}|^2 + x^{n}_{j}|\psi^{n}_{j}|^2 \right) \, . \label{numerov}
\end{eqnarray}
This algorithm serves to integrate second order ordinary differential equations, and is locally sixth order accurate (see \cite{cphysics} for details). As we are solving the evolution equation (\ref{schroedinger}) with a second order accurate differencing algorithm, the fact that we are solving the constraint (\ref{poisson}) with a more accurate algorithm does not mean that the evolution itself should be better than second order.

\subsection{Boundary conditions}
\label{subsec:bc}
Eq.~(\ref{numerov}) is to be integrated inwards and thus it is necessary to fix the value of the gravitational potential at the two outermost points of the numerical grid, i.e., we should deal now with boundary conditions. The easiest boundary condition for the gravitational potential is 
\begin{subequations}
\label{bcU}
\begin{eqnarray}
U(x_{N-1}) &=& -M(x_{N-1})/x_{N-1} \, , \\
U(x_N) &=& -M(x_N)/x_N \, .
\end{eqnarray}
\end{subequations}
However, this boundary condition has to be taken carefully. Strictly speaking, (\ref{bcU}) is only valid and consistent if the total mass is confined to the region $x < x_{N-1}$ (or equivalently, $M(x_{N-1})=M(x_N)$), since Eq.~(\ref{bcU}) is the form the gravitational potential takes in vacuum. 

In consequence, expression (\ref{bcU}) is equivalent to impose the condition $|\psi(\tau, x_N)| \rightarrow 0$ on the wave function. This is, for example, the shooting condition to find equilibrium configurations we applied in Sec.~\ref{subsec:neq}. Thus, we are forcing the system not to leave ever the computational domain, and then there cannot be outgoing waves at all: the outer boundary is a perfect reflective wall.

The boundary condition is an important issue since there is no chance to apply the Sommerfeld boundary condition on the wave function $\psi$ as in the relativistic case\cite{seidel90,phi2}, because in the present situation the evolution equation is not hyperbolic. Moreover, if the boundary condition~(\ref{bcU}) were applied alone, it would only work for the equilibrium configurations -for which the number of particles has to be preserved- but the evolution of other arbitrary systems, including those that eject matter, would be incorrect. 

In order to avoid this, we implemented a \textit{sponge} over the outermost points of the grid, which consists in adding an imaginary potential $V(x)$ to the Schr\"odinger equation; the expression we use is
\begin{equation}
V_j = -\frac{i}{2} V_0 \left\{ 2 + \tanh \left[(x_j-x_c)/\delta \right] - \tanh \left( x_c/\delta \right) \right\} \, , \label{imagpot}
\end{equation}

\noindent which is a smooth version of a step function with amplitude $V_0$ and width $\delta$. We choose this shape for the imaginary potential because in the absence of gravity, the wave function decays exponentially (see Sec.~\ref{toymodel} below), and the smoothness is to diminish the effects of our finite differencing schemes.

Notice that the definition of an imaginary potential makes physical sense and fits well into the SN system. This is seen if we recalculate the conservation of probability which now reads (see Eq.~(\ref{current_conservation}))
\begin{equation}
\partial_{\tau} \rho + \frac{i}{2x} \left[ \psi \partial^2_x (x \psi^\ast) - \psi^\ast \partial^2_x(x\psi) \right] = -2i V \rho \, .\label{conservation_im}
\end{equation}
The source term on the r.h.s. is always proportional to the density of particles, and the minus sign warranties the decay of the number of particles at the outer parts of our integration domain, that is, the imaginary potential behaves as a sink of particles. 

One last remark. Our original SN system is not physically related to the sponge region. Then, we adjust the parameters of the sponge so that it only covers some points at the outer part of our grid; actually, we have fixed the spatial range of the sponge to $\Delta_{\textrm{sponge}}= 2(x_N-x_c)$. The values of $\Delta_{\textrm{sponge}}$ and $x_N$ are fixed by hand and then we calculate $x_c$ and set $\delta= \Delta_{\textrm{sponge}}/10$. Notice too that $V_N=-iV_0$. In this manner, we shall call \textit{physical region} to the grid points that accomplish $x_j \leq (x_N-\Delta_{\textrm{sponge}})$, and this will be our region of physical interest.

\subsubsection{Toy model: Imaginary 1-D square well potential}
\label{toymodel}
However, the sponge~(\ref{imagpot}) is not a perfect absorber, and to see how it works in reality we consider the following toy model (a more general discussion can be found in\cite{israeli}). 

We simplify the system and consider that at the last points of our numerical grid, where the sponge is implemented, the gravitational potential can be neglected. Then, our physical system at the last points would be similar to the case in which a Schr\"odinger wave function is scattered back off by an imaginary square well potential of the form $V(0 \leq x \leq x_c)=-iV_0$ and $V(x)=0$ elsewhere. 

As in usual quantum mechanics examples, the radial Schr\"odinger wave function $x\psi(\tau,x)=u_r(x)e^{-iE\tau}$ has the form
\begin{equation}
u_r (\tilde{x})= \left\{
\begin{array}{ccc}
e^{ik\tilde{x}} + R e^{-ik\tilde{x}} \, &,&  \tilde{x}  < 0 \, , \\
Pe^{ik^\prime \tilde{x}} + Q e^{-ik^\prime \tilde{x}} \, &,& 0 \leq \tilde{x} \leq \ell \, , \\
0 \, &,& \ell < \tilde{x} \, ,
\end{array}
\right.
\end{equation}
where $k=\sqrt{E /V_0}$ and ${k^\prime}^2=i+k^2$ are the wave numbers outside and inside the well potential, respectively. The presence of imaginary number $i$ reveals the imaginary nature of the well potential. 

The coefficients $R$, $P$, and $Q$ are already normalized with respect to the incident wave $e^{ikx}$, while the spatial coordinate is normalized in the form $\tilde{x}=x\sqrt{2V_0}$ and then $\ell=\Delta_\textrm{sponge} \sqrt{2V_0}$. Notice that we have added the condition of perfect reflection at the right boundary $\tilde{x}=\ell$, as it is the case for our numerical system. 

Using the continuity and differentiability conditions for the wave function at the boundaries $\tilde{x}=0$ and $\tilde{x}=\ell$, we find the following expression for the reflection coefficient
\begin{equation}
|R|^2 = \left| \frac{ik \sin (k^\prime \ell)+k^\prime \cos (k^\prime L)}{ik 
\sin (k^\prime \ell)-k^\prime \cos (k^\prime L)} \right|^2 \, . \label{reflex}
\end{equation}
This formula is also valid for a {\it real} well potential under the same boundary conditions, and for such case it gives the obvious solution $|R|^2=1$.

\begin{figure}[htp]
\includegraphics[width=6cm]{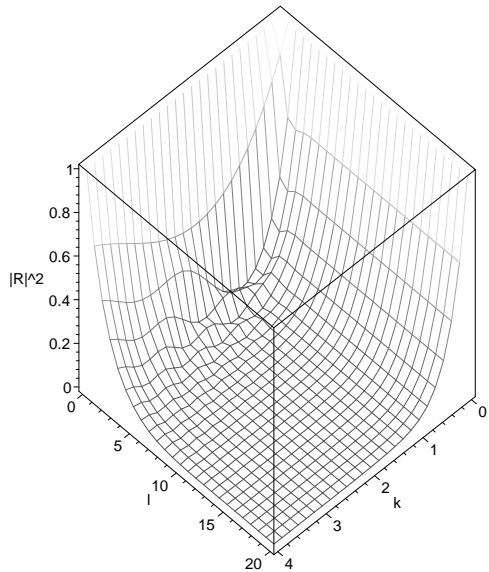}
\caption{\label{fig:xxx}
The reflection coefficient $|R|^2$, 
Eq.~(\ref{reflex}), as a function of $k$ and $\ell$, for the linear toy model shown in text. The reflected number of particles is suppressed for large values of both parameters $k,\ell$.} 
\end{figure}

A graph of the reflection coefficient $|R|^2$ as a function of $(k,\ell)$ is shown in Fig.~\ref{fig:xxx}, where we see that $|R|^2$ is highly suppressed for $k \gg 1$ and $\ell \gg 1$. The former condition means that more modes can get into the sponge, while the latter one means that the sponge region is sufficiently large as to let the modes die out inside the sponge.

Going back to our original system, it is then obvious that low energy modes will always be reflected in some amount and they will always contaminate what we called the physical region. However, the reflected amount of matter is not as large as in the toy model above, since the imaginary potential~(\ref{imagpot}) is a smooth function. Then, the depth and the size of the imaginary potential can be adjusted to let most of the modes enter the sponge. 

Fortunately, because of the fitness of the imaginary potential into the SN system, the sponge region could be larger than the region of physical interest without resulting in an erroneous evolution of the SN system. The appropriateness of the sponge is analyzed in Sec.~\ref{boundary}.

\subsection{Accuracy}
In the relativistic case, the EKG system is redundant and one of the Einstein equations can be used to measure how accurate the numerical solution is. For instance, in\cite{phi2}, the $\{t,t\}$ and $\{r,r\}$ components of the Einstein equations were solved and then it was determined whether the $\{t,r\}$ component was accurately satisfied. 

Following the relativistic case, Eq.~(\ref{conservation_im}) gives a criterion to measure the accuracy of our numerical code for the complete grid of integration, since it is an equation our system should accomplish independently of the potential involved. As said before, such equation is the weak-field constraint equation of the relativistic system. Then, we define our accuracy parameter to be:
\begin{equation}
\beta := \partial_\tau \rho + \frac{i}{2x} \left[ \psi \partial^2_x (x \psi^\ast) - \psi^\ast \partial^2_x(x\psi) \right] + 2i V \rho \, . \label{beta}
\end{equation}
In the continuum limit, $\beta \equiv 0$ holds for an exact solution, thus the magnitude of this quantity will tell us how close our numerical solution is to the exact one. 

However, it should be noticed that our accuracy parameter obeys the scaling 
transformation $\beta = \lambda^6 \hat{\beta}$. So, the value of $\beta$ 
should be interpreted carefully otherwise we could claim that the most 
accurate physical solution is that with the smallest value of $\lambda$, which 
cannot be true (accuracy cannot be just a matter of scaling).

In most cases (see for instance Sec.~\ref{scalinrels}), it would be more useful to define a $\lambda$-independent parameter. A simple and straightforward definition we can think of is a \textit{relative error} parameter of the form
\begin{equation}
\Delta \beta = \frac{||\sum_i \beta_i||_2}{\sum_i ||\beta_i||_2} \, . \label{relbeta}
\end{equation}
Here, $\beta_i$ is each of the terms that appear in the momentum constraint, whether relativistic or non-relativistic. For instance, $\beta_i$ represents each of the terms we need to calculate our accuracy parameter $\beta$ in Eq.~(\ref{beta}); see\cite{phi2} for the definition of a $\beta$ for a relativistic system. Clearly, $\Delta \hat{\beta}=\Delta \beta$ for the SN system, and we will indicate its value whenever the interpretation of $\hat{\beta}$ is confusing.

As a final note, we should mention that the accuracy of the evolutions depends on the ratio $\Delta \tau / (\Delta x)^2$, the latter should be less than unity in order to have a reliable evolution of the Schr\"odinger equation, see\cite{nr}. However, it should be noticed here that, according to the scale transformation~(\ref{scale1}), $\Delta \tau / (\Delta x)^2 = \Delta \hat{\tau} / (\Delta \hat{x})^2$, and then the computational effort is the same for the original configuration as for the properly sized one. Actually, the condition $\Delta \tau / (\Delta x)^2 < 1$ is a very restrictive one, and the responsible for the very large runs we need in order to get a reliable evolution at late times.

\section{Results}
\label{results}
In this section, we show representative runs of the different issues discussed in the previous sections. The numerical method to evolve the SN system is tested thoroughly about the effect of the boundary conditions, accuracy and convergence of the numerical results. In brief, the tests show that the numerical results are reliable even in long runs, as far as the boundary conditions are set up appropriately.

Another test considered is the evolution of $0$--node equilibrium configurations and their perturbations. The numerical evolution reproduces the semianalytical results of the previous sections and shows that $0$--node equilibrium configurations are intrinsically stable. On the other hand, it is shown that the numerical method preserves the scaling properties of the SN system even in the cases of non--equilibrium solutions. Also and for completeness, the equivalence of the (relativistic) EKG and the SN systems is studied in the weak field limit and found to be correct within the numerical accuracy. Finally, excited equilibrium configurations are found to be intrinsically unstable and that they all decay to $0$--node solutions.

\subsection{Boundary effects}
\label{boundary}
Being the boundary conditions an issue of particular importance in this work, it is important to determine the reliability of the sponge for the purposes of this paper, since part of the initial mass could be ejected to infinity and forced to interact with the imaginary potential at the outer boundary.

The form in which we test our boundary condition is as follows. The \textit{ideal} numerical evolution is that in which the numerical boundary is very far away from the physical boundary, $x_p \ll x_N$, so that no scalar matter has reached $x_N$ for the time range the numerical evolution is followed. In this manner, we can assure that the solution inside the physical region $0 \leq x\leq x_p$ is the correct one.

On the other hand, let us suppose that $x_p \sim x_N$. and that a sponge is implemented on the region $x_p < x \leq x_N$. If the numerical evolution in the range $0 \leq x\leq x_p$ coincides with that of the ideal case above, then we say that an appropriate sponge was implemented.

\begin{figure}[htp]
\includegraphics[width=8cm]{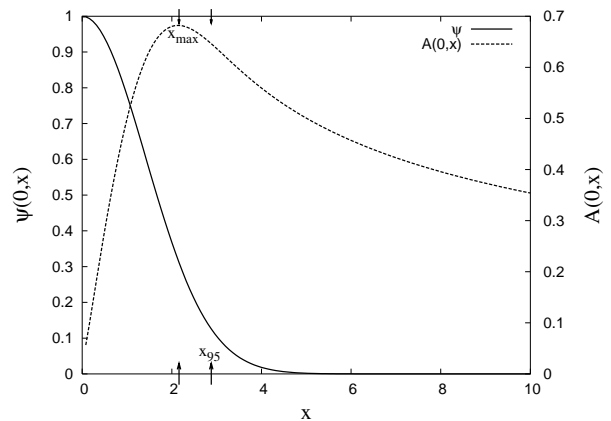}
\caption{\label{fig:gaussini} 
Initial profiles of the wave function 
$\psi(0,x)=\exp[-(x/2)^2]$ (left axis) and the metric function $A(0,x)=M(0,x)/x$ (right axis) of \textit{Run I}. $x_{\textrm{max}}$ indicates the position of the maximum value of $A(0,x)$, and $x_{95}$ is the 95\% mass radius.} 
\end{figure}

With this in mind, we have numerically evolved the initial Gaussian profile shown in Fig.~\ref{fig:gaussini}, which is of the form $\psi(0,x)=\exp[-(x/2)^2]$; also shown is the corresponding metric function $A(0,x)$, see Eq.~(\ref{grr_expansion}). We have marked the radii $x_{95}$ (the so called 95\% mass radius) and $x_{\textrm{max}}$, where the latter indicates the radius at which $A(t,x)$ reaches its maximum value. In the relativistic systems, $x_{\textrm{max}}$ has been a good quantity to follow the evolution of scalar systems, see\cite{seidel91,phi2}.

The time and space resolutions of the numerical evolution were, respectively, $\Delta \tau=3\times 10^{-3}$ and $\Delta x = 0.08$. Also, the numerical and physical boundaries were $x_N=960$ and $x_p=40$, respectively, and no sponge was implemented. We shall call this case \textit{Run I}. 

\begin{figure}[htp]
\includegraphics[width=8cm]{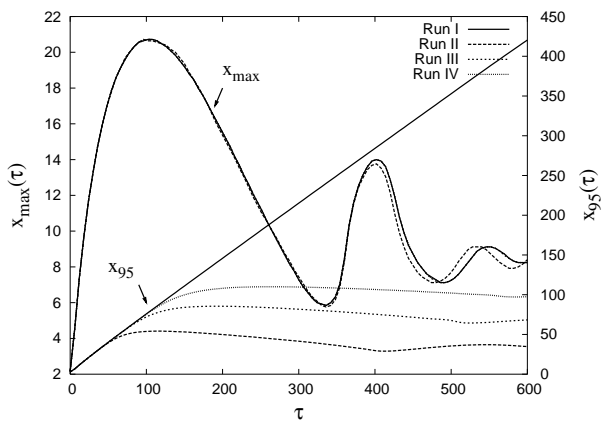}
\caption{\label{fig:gaussrNa}Evolution of the two radii $x_{\textrm{max}}$ (left axis) and $x_{95}$ (right axis) for \textit{Runs I}, \textit{II}, \textit{III} and \textit{IV}; see text for details.}
\end{figure}

In Fig.~\ref{fig:gaussrNa} we show the evolution of $x_{\textrm{max}}$ and $x_{95}$ for \textit{Run I}. We notice that $x_{\textrm{max}}$ oscillates and approaches to a finite fixed value, while $x_{95}$ grows almost linearly for all the time the evolution was followed. 

\begin{figure}[htp]
\includegraphics[width=8cm]{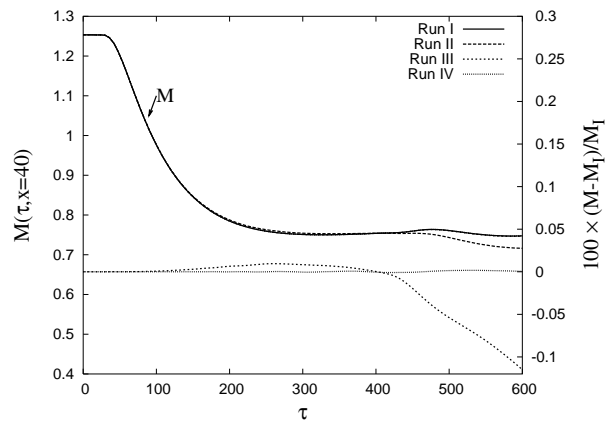}
\caption{\label{fig:gaussrNb}Mass number $M(\tau,x=40)$ (left axis) for all \textit{Runs I} to \textit{IV}. Also shown is the relative mass difference (right axis) of \textit{Runs III} and \textit{IV} with respect to \textit{Run I}. For \textit{Run IV}, the mass deviation is almost unnoticeable; see text for details.} 
\end{figure}

Roughly speaking, $x_{\textrm{max}}$ gives us an estimate of the radius at which most of the gravitational interaction is contained in, while $x_{95}$ can be seen as a tracer of the ejected mass that escapes to infinity. Thus, we can say that a finite sized configuration is formed, and that the ejected scalar matter is far from reaching the numerical boundary. That is, the solid curves in Fig.~\ref{fig:gaussrNa} and~\ref{fig:gaussrNb} are what we would see in the ideal numerical evolution.

Shown in Fig.~\ref{fig:gaussrNb} is the mass contained inside the physical region $0\leq x \leq 40$ as the evolution proceeds. For the case of \textit{Run I} (solid curves), we notice that part of the mass leaves the physical region ($\tau \sim 300$) but a little bit of it returns ($\tau \sim 500$) and leaves again ($\tau \sim 600$). This return of scalar matter cannot be attributed to reflection at $x_N$ since no matter has reached the numerical boundary. But, this effect is due to matter that leaves the physical region with a velocity less than the escape velocity of the system.

Being \textit{Run I} an example free of noise from the numerical boundary, we will compare it with other runs in order to test the reliability of a sponge as a boundary condition. For sake of simplicity, the width of the imaginary potential was fixed to $\delta = (x_N-x_p)/10$ in the runs described next.

\textit{Run II} has the same time and space resolutions as \textit{Run I}, but the values of the physical and numerical boundaries are, respectively, $x_p=40$ and $x_N=240$, and the sponge depth (see Eq.~(\ref{imagpot})) is $V_0=55.0$. For this case, $x_{95}$ reaches a maximum value but never reaches the value of the numerical boundary; this means that the ejected matter leaves the physical boundary  (from $\tau \sim 100$ on) and is absorbed by the sponge. 

On the other hand, the value of $x_{\textrm{max}}$ (see Fig.~\ref{fig:gaussrNa}) again oscillates and approaches to a finite value, but we note a shift in the oscillations with respect to \textit{Run I}. The reason for this can be found if we look at the plot of the mass number $M(\tau,x=40)$ in Fig.~\ref{fig:gaussrNb}.

 First of all, the reflection of matter is noticeable around $\tau \sim 200$, which is due to the sharpness of the sponge, recall that $V_0=55.0$. Also, the size of the physical region is very small, since the sponge absorbs matter that should have returned to the physical region because it did not have the required velocity to escape to infinity, as it is shown by \textit{Run I}. At the end of the run, there is a mass difference of about $5\%$ between \textit{Run I} and \textit{Run II} in Fig.~\ref{fig:gaussrNb}, which is sufficient to make \textit{Run II} evolve apart from \textit{Run I} at late times.

\textit{Run III} uses the same parameters as \textit{Run II}, but now $V_0=1.0$. In general, the results are qualitatively the same as for \textit{Run II}, but the amounts of reflected and absorbed matter are drastically reduced now that the sponge is smoother. As seen in the corresponding $x_{95}$ in Fig.~\ref{fig:gaussrNa}, the ejected matter can travel a longer distance inside the sponge region and some of it can return to the physical region without being completely annihilated. The mass difference with respect to \textit{Run I} at the end of the run is around $0.1\%$ (see Fig.~\ref{fig:gaussrNb}), and then the oscillations of $x_\textrm{max}$ are not shifted noticeably with respect to those of \textit{Run I}.

\textit{Run IV} has the same parameters as \textit{Run III}, except for the physical and numerical boundaries now at $x_p=80$ and at $x_N=280$, respectively. That is, the sponge region has the same size in \textit{Runs II,III} and \textit{IV}. The latter is the most similar to \textit{Run I}, being the mass deviation in Fig.~\ref{fig:gaussrNb} less than $0.001\%$ at the end of the simulation. This is because the physical region is larger than in \textit{Run III}, and then more ejected matter can return to the region $x < 40$ without being affected by the sponge.

All the results presented here show that with an appropriate sponge we can obtain the same results as if the numerical boundary were very far away, with the advantage that \textit{Run IV} needs less numerical efforts than \textit{Run I}\footnote{The runs were followed up to $\hat{\tau}=600$, but recall that the physical dimensionless time is such that $\tau > 10^3 \hat{\tau}$; thus, as far as we know, the runs in this section are the largest reported in the literature.}.

\begin{figure}[htp]
\includegraphics[width=8cm]{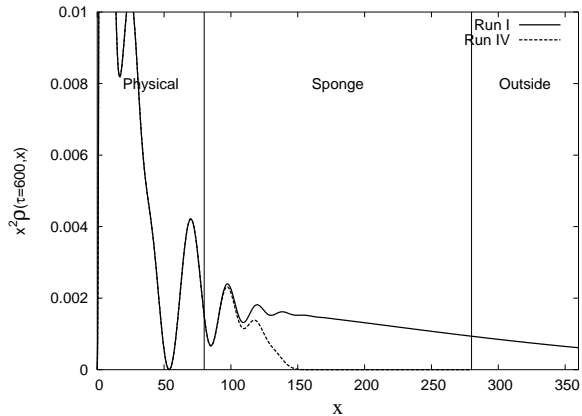}
\caption{\label{fig:grxx}
Comparison of the final profiles of $x^2 \rho(\tau=600,x)$ corresponding to \textit{Run I} and \textit{Run IV}, where we have marked the physical and sponge regions for the latter. The profiles are quite similar inside the physical region, but the profile of \textit{Run IV} rapidly vanishes inside the sponge region, and then we are accomplishing the boundary condition $|\psi(\tau,x_N)|\rightarrow 0$.} 
\end{figure}

Finally, we show in Fig.~\ref{fig:grxx} a comparison of the final profiles of $x^2 \rho$ corresponding to \textit{Run I} and \textit{Run IV}. As it should be for a good boundary condition, the profiles are quite similar inside the physical region, which is our region of interest. Moreover, it is seen that the expected behavior $|\psi (\tau,x_N)| \rightarrow 0$ is achieved, which is the boundary condition compatible with Eq.~(\ref{bcU}).

In conclusion, the implementation of a good sponge should follow, in general, the indications found in Sec.~\ref{toymodel}: a smooth and large sponge will make it well. However, the examples presented in this section gives more details about how smooth and how large a sponge could be in order to have a reliable run and a low numerical effort. Also, we learned that a good sponge should be accompanied by a sufficiently large physical region.

\subsection{Accuracy and Convergence}
\label{convergence}
Once we have shown the reliability of the boundary conditions, we can perform accuracy and convergence tests. 

\begin{figure}[htp]
\includegraphics[width=8cm]{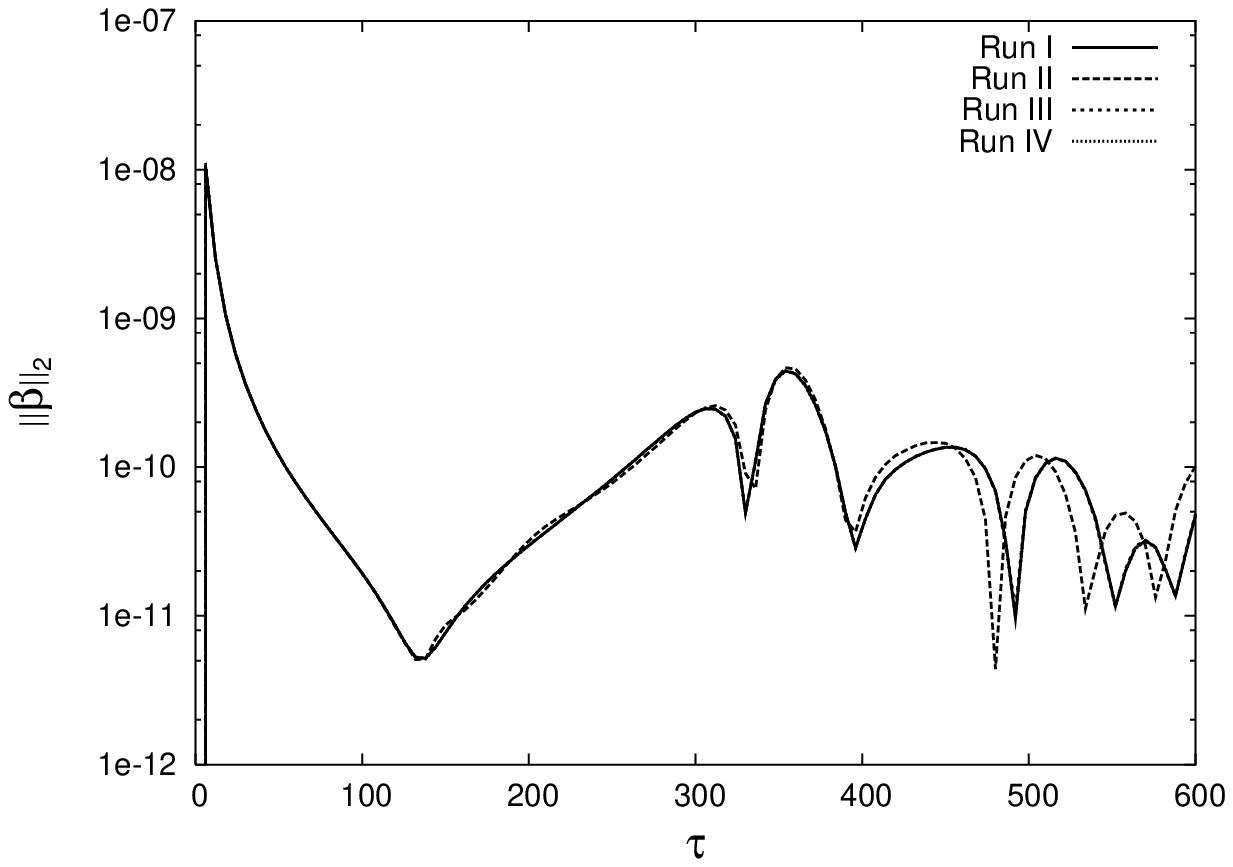}
\includegraphics[width=8cm]{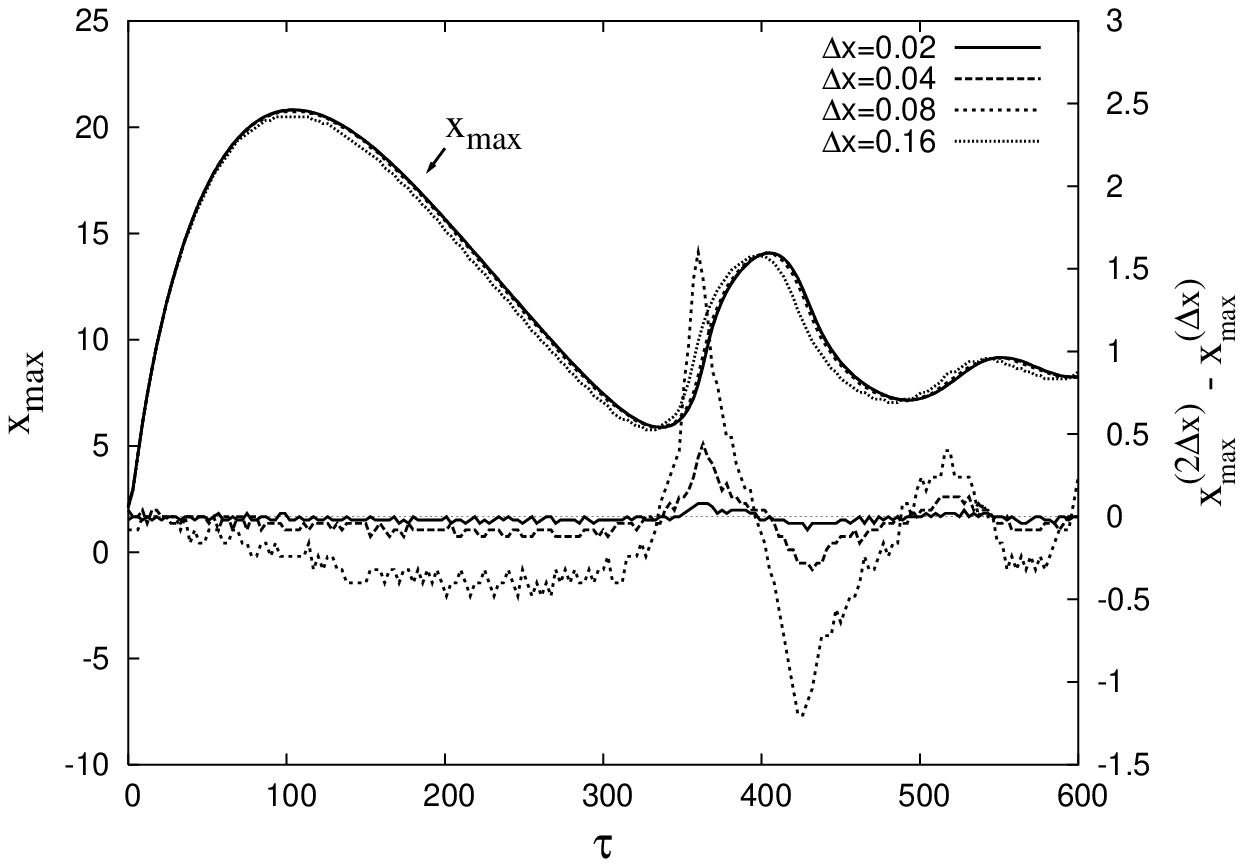}
\caption{\label{fig:gbeta} (Top) The value of $||\beta||_2$ for \textit{Runs I, II, III and IV}. In all these cases, $\Delta \beta < 10^{-6}$. (Bottom) Evolution of $x_\textrm{max}$ (left vertical axes) for the same case as in \textit{Run IV}, but with a fixed time step $\Delta \tau=2\times 10^{-4}$ and four different spatial resolutions $\Delta x=0.02$, $0.04$, $0.08$ and $0.16$. It can be seen that the numerical solution converges as $\Delta x \rightarrow 0$ in phase and amplitude. Also shown (right vertical axes) is the difference in $x_\mathrm{max}$ between runs with spatial resolutions $\Delta x$ and $2\Delta x$; the difference is four times smaller if the spatial resolution is doubled, see text below.}
\end{figure}

To begin with, we show in Fig.~\ref{fig:gbeta} how $||\beta||_2$ evolves in time; its value is less than $10^{-8}$ for \textit{Runs I}, \textit{II}, \textit{III} and \textit{IV}. The values of $||\beta||_2$ are the same because the ratio $\Delta \tau /(\Delta x)^2$ is also the same for all runs. It is clear here that all deviations from \textit{Run I} are due to matter reflected by the sponge. 

Another test we consider important is that of convergence, that is, whether the solutions of our numerical scheme approach the exact solution as we increase the spatial resolution. To investigate this, we perform four different runs for the same system as in \textit{Run IV} with a fixed time step $\Delta \tau=2\times 10^{-4}$, and four spatial resolutions $\Delta x=0.02$, $0.04$, $0.08$ and $0.16$; the corresponding evolutions of $x_\textrm{max}$ are shown in Fig.~\ref{fig:gbeta}.

Only the coarsest run ($\Delta x=0.16$) deviates to second order in phase from the other three, and it was found that such coarse resolution is out of the convergence regime after certain finite time of evolution; such deviation should be attributed to the discretization used and not to reflected matter. Moreover, it can be noticed that the solutions indeed converge to the solid line as the spatial resolution is increased.

We also show the deviations when we compare the runs by pairs, $f^{(2\Delta x)}-f^{(\Delta x)}$, where $f^{(\Delta x)}$ is any function of our system that was solved using a fixed spatial resolution $\Delta$. For the case of $x_\textrm{max}$, we notice in Fig.~\ref{fig:gbeta} that the deviations are four times smaller when the spatial resolution is doubled. 

\subsection{Equilibrium configurations in the ground state}
Another important test is provided by the existence of stationary solutions of the SN system of equations which were found in Section~\ref{subsec:neq}. In short, these solutions are obtained by assuming that the time dependence of the Schr\"odinger field is of the form $\psi = e^{i \gamma \tau} \phi(r)$, which in turn implies that both the gravitational potential $U(x)$ and the probability density $\rho=|\psi|^2$ are time-independent.
 
\begin{figure}[htp]
\includegraphics[width=8cm]{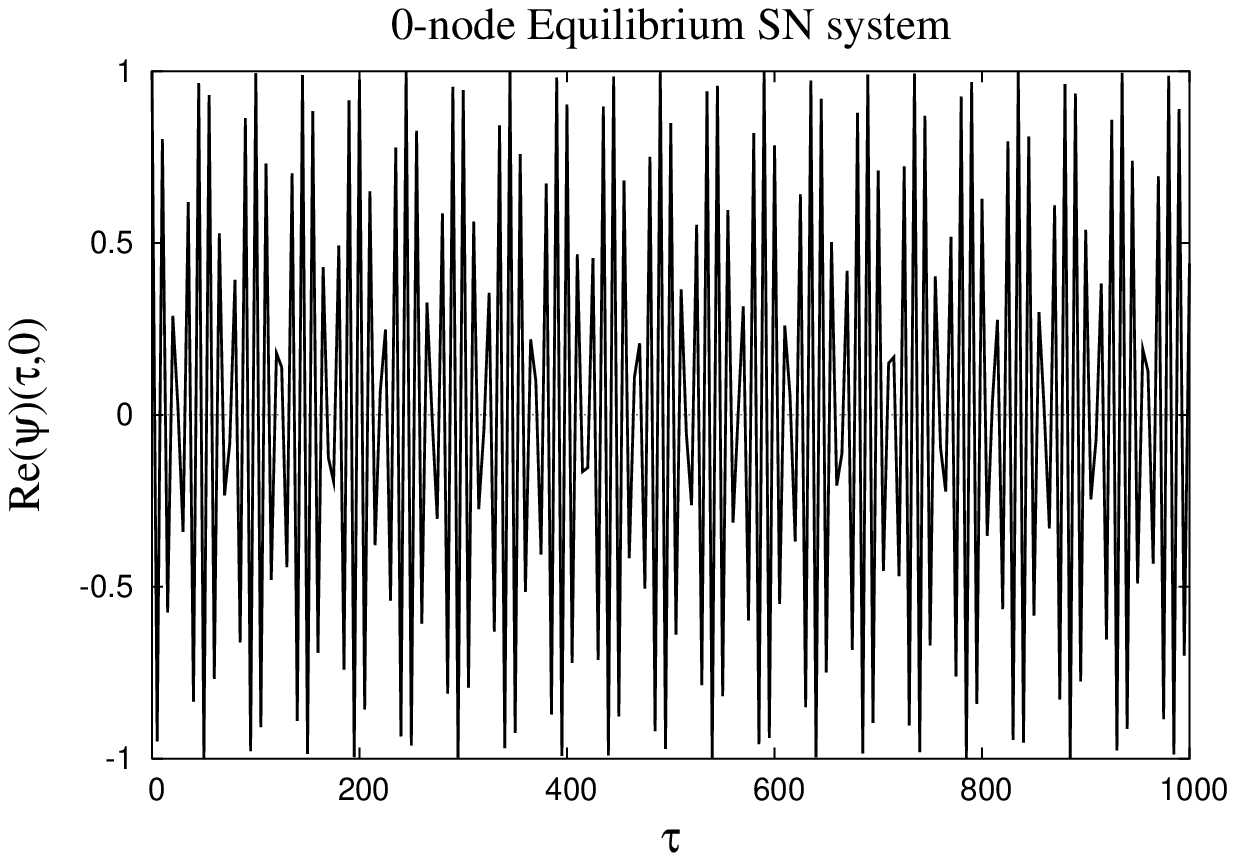}
\includegraphics[width=8cm]{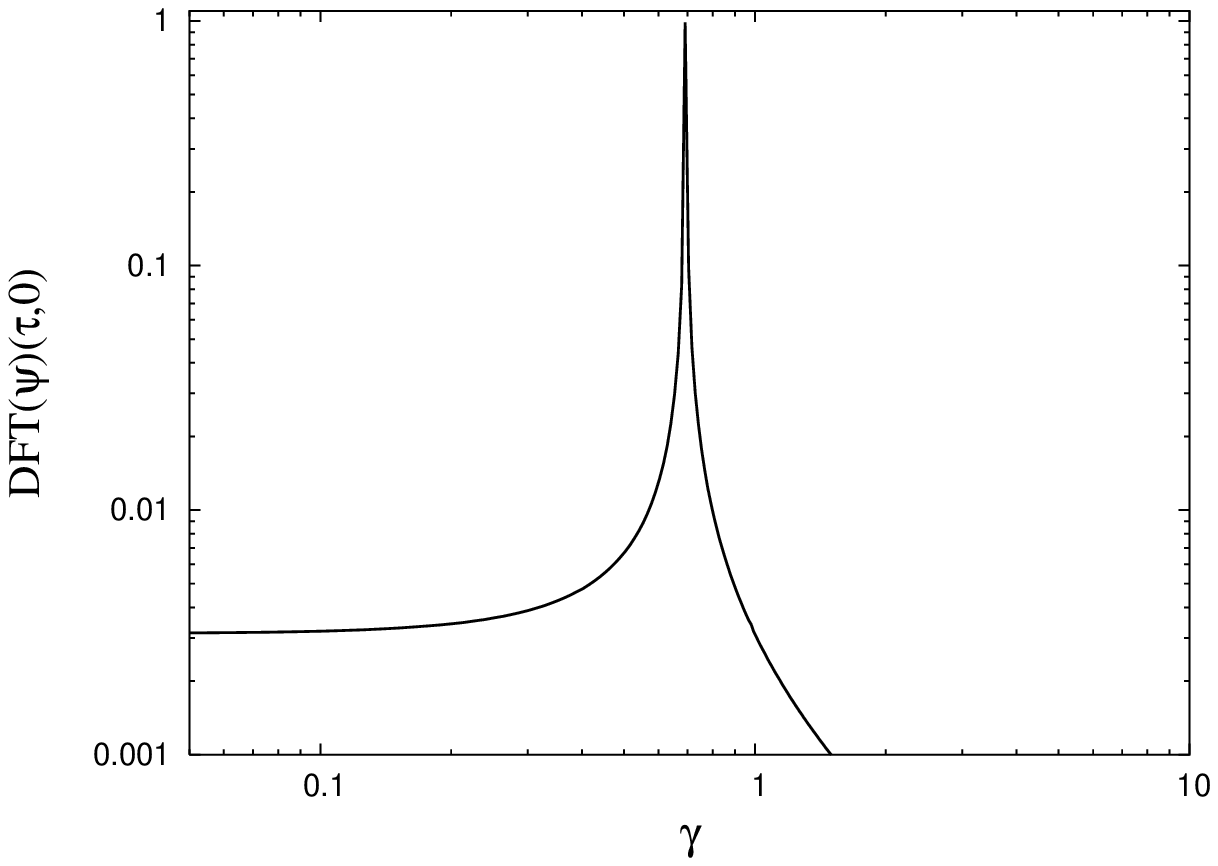}
\caption{\label{fig:test1} (Top) The evolution of ${\rm Re}(\psi(\tau,0))$ for a 0--node equilibrium configuration. (Bottom) The DFT of the evolved solution shows a unique harmonic mode for an angular frequency of $\gamma=0.697$, which coincides well with the eigenvalue problem of Section~\ref{subsec:neq}.}
\end{figure}

In Fig.~\ref{fig:test1}, it is shown the evolution of ${\rm Re}(\psi(\tau,0))$ for a 0--node equilibrium configuration. The boundaries were located at $x=50$ with a time resolution of $\Delta t = 0.001$ and a spatial resolution of $\Delta x = 0.02$. It is observed that the wave function oscillates harmonically, as its Fourier transform (FT) shows a unique harmonic mode with an angular frequency $\gamma=0.697$, in good agreement with the eigenvalue solution found in 
Section~\ref{subsec:neq}.\\

As was discussed in Section~\ref{subsec:bc}, the boundary condition~(\ref{bcU}) imposed on the gravitational potential makes the energy density vanish at the outer boundary. As there is not any violent collapse or explosion in this system, such boundary condition is appropriate for $0$--node equilibrium configurations. Fig.~\ref{fig:test1} also shows that the numerical code is stable and, within numerical precision, reproduces the expected analytical results.\\

\subsubsection{Observing extra modes}
\begin{figure}[htp]
\includegraphics[width=8cm]{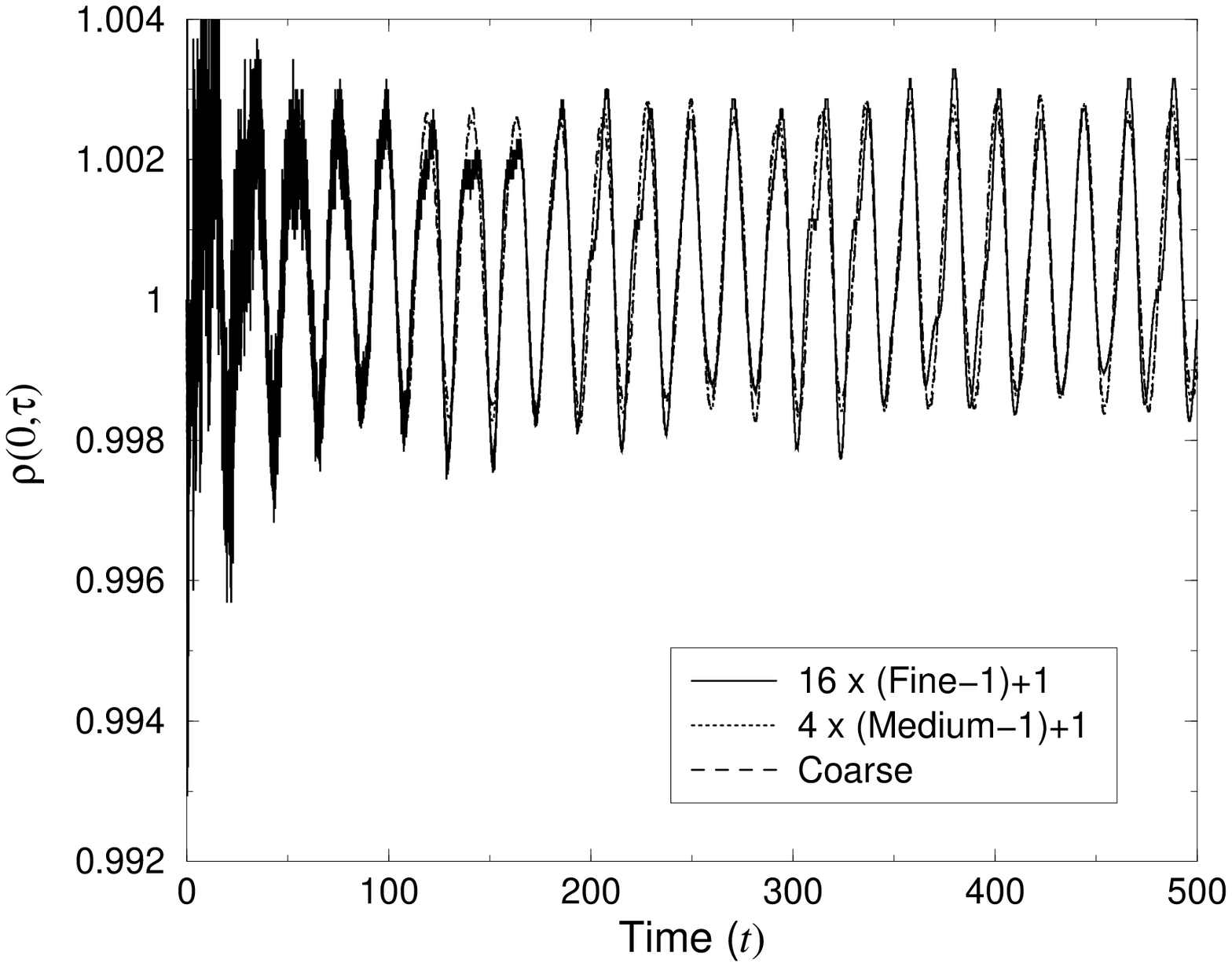}
\includegraphics[width=8cm]{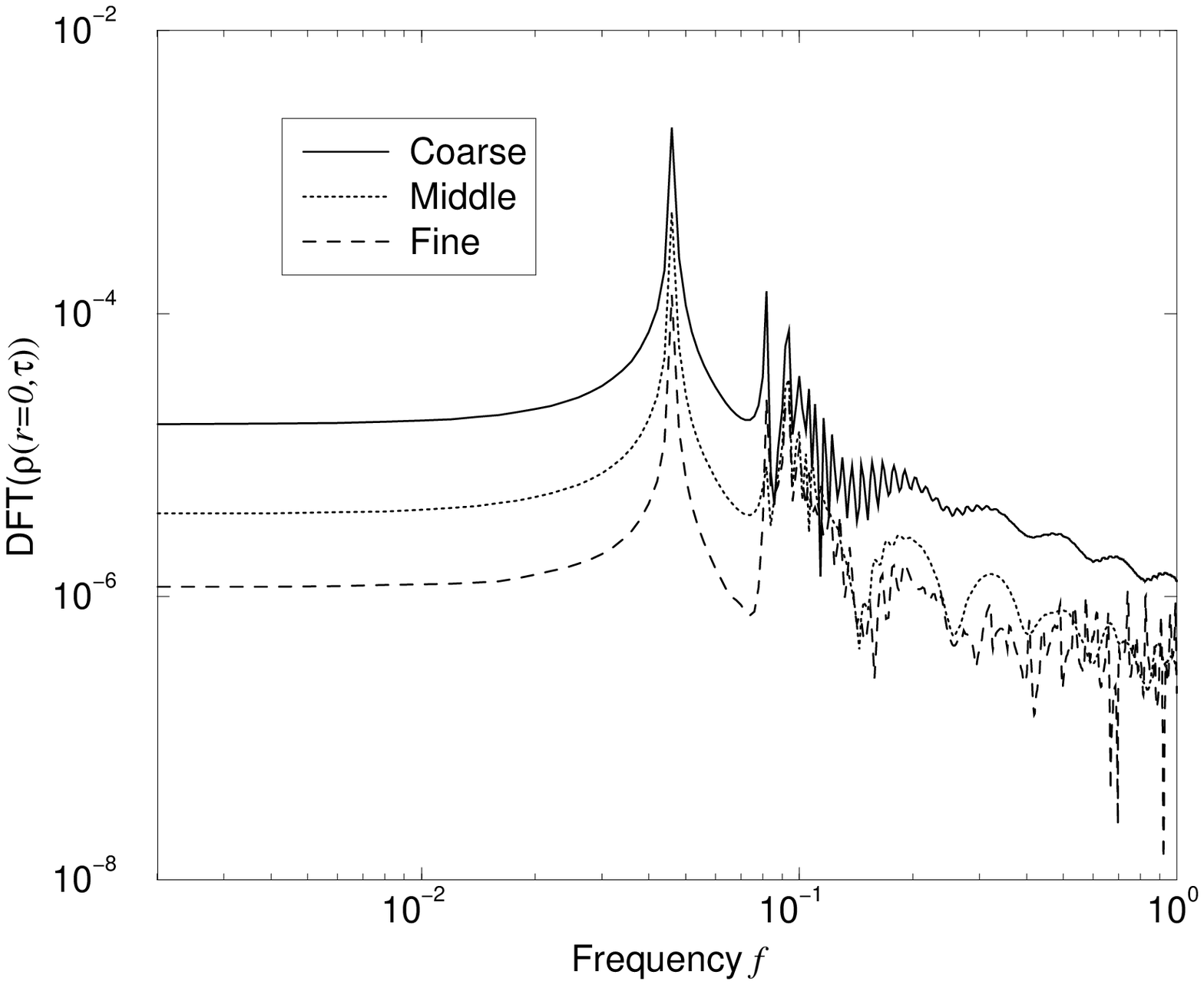}
\caption{\label{fig:DFT_N0_eq} 
(Left) The central value $\rho (\tau,0)$ 
at different resolutions $\Delta x=0.04, \, 0.08, \, 0.16$ using a fixed value of $\Delta t = 10^{-3}$. The plots show that the numerical code is second order convergent. It can be observed that for the finest resolution there is an initial noise due to the fact that the quantity $\Delta t/(\Delta x){}^2$ is bigger than in the other two cases. (Right) The FT of $\rho (\tau,0)$ shows the first peak which we associate to the quasinormal frequency at $f\simeq 0.046$, see Sec.~\ref{subsec:perturbations}.} 
\end{figure}

The evolution of the $0$--node configuration was observed carefully in order to look for the modes predicted by the perturbation theory described in Sec.~\ref{subsec:perturbations}. For this, we searched for the perturbations of the energy density as they can be easily isolated, see Eqs.~(\ref{perturbations}). We show in Fig.~\ref{fig:DFT_N0_eq} the oscillations on the energy density of the system due to the perturbations introduced by the discretization of our grid. Actually, we are comparing three runs with three different resolutions $\Delta x=0.04, \, 0.08, \, 0.16$ with the time resolution fixed at $\Delta \tau = 10^{-3}$.
  
It is noticed that the perturbations with the smallest amplitude corresponds to the run with the finest spatial resolution, as it can be seen from the FT of the oscillations also shown in Fig.~\ref{fig:DFT_N0_eq}. This plot also indicates that the oscillation frequency $f \simeq 0.046$ is the same despite the resolution used, which coincides with the value found by solving the perturbation equations for the 0--node configuration in Section~\ref{subsec:perturbations}. Last but not least, the amplitude of the perturbations also show that the numerical code is second order convergent: for a fixed $\Delta \tau$, the amplitude of the perturbations are four times smaller if we double the spatial resolution, in concord with the results of Sec.~\ref{convergence}.

\begin{figure}[htp]
\includegraphics[width=8cm]{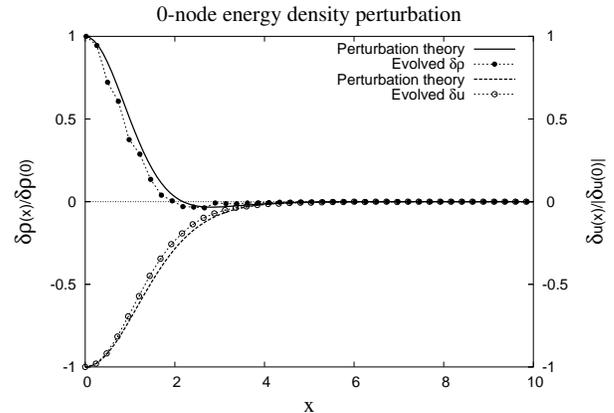}
\caption{\label{fig:perts_ev} The normalized perturbation profiles of the energy density (left axis) and the gravitational potential (right axis) of a 0--node equilibrium configurations. Shown here is a comparison between the evolved profile shown in Figs.~\ref{fig:test1} and~\ref{fig:DFT_N0_eq}, and the solutions obtained in Section~\ref{subsec:perturbations}. Although there is not a complete correspondence, the similarity of the graphs suggests that we indeed see the same kind of perturbation in both cases. The spatial resolution of the evolved profile was larger than shown in the plot.} 
\end{figure}

For further comparison, we show in Fig.~\ref{fig:perts_ev} the profiles of the perturbations as obtained from the evolved systems and the perturbation equations~(\ref{perturbed_ivp}). Although the correspondence is not exact, the results still suggest that the perturbations in Figs.~\ref{fig:perturbations} and~\ref{fig:DFT_N0_eq} are of the same kind. 

A general conclusion of this section is that, as anticipated from the perturbations analysis, $0$--node equilibrium configurations are intrinsically stable under small radial perturbations.

\subsection{Scaling relations}
\label{scalinrels}
For the scale invariance~(\ref{scale1}) to be really useful, we have to be sure that the initial scaling relation is preserved by the numerical code during the evolution of the SN system. Also, we would like to test the post--Newtonian approximation made for the relativistic system, i.e., whether the SN system is the weak field limit of the EKG equations. That the latter is true has been shown before in the case of boson stars\cite{seidel91}, but we want to show it also for the case of oscillatons.

To test both features above in one stroke, we show the numerical evolution of a $0$-node Newtonian oscillaton which was perturbed so that its mass was increased roughly by $20$\%; such initial profile was used to feed the relativistic numerical scheme used in\cite{phi2}, following Eqs.~(\ref{newtoscillaton}). 

The scale invariant quantities we used for the SN system were $\Delta \hat{x}=0.04$, $\hat{\tau}=3.0 \times 10^{-4}$, with the physical and numerical boundaries at $\hat{x}_p=80$ and at $\hat{x}_N=160$, respectively. The numerical evolution of the SN system was followed up to a time $\hat{\tau}=27$. 

On the other hand, to set up the initial profile for the relativistic case, the EKG system, we take the scaling parameter $\lambda=0.032$. Because of the scale transformation, the spatial and time resolutions used were $\Delta x=1.25$ and $\Delta \tau = 0.0125$, respectively, while the physical boundary was fixed at $x_p=2500$. The relativistic run was followed up to $\tau = 25000$, that is, just $10$ crossing times.

\begin{figure}[htp]
\includegraphics[width=8cm]{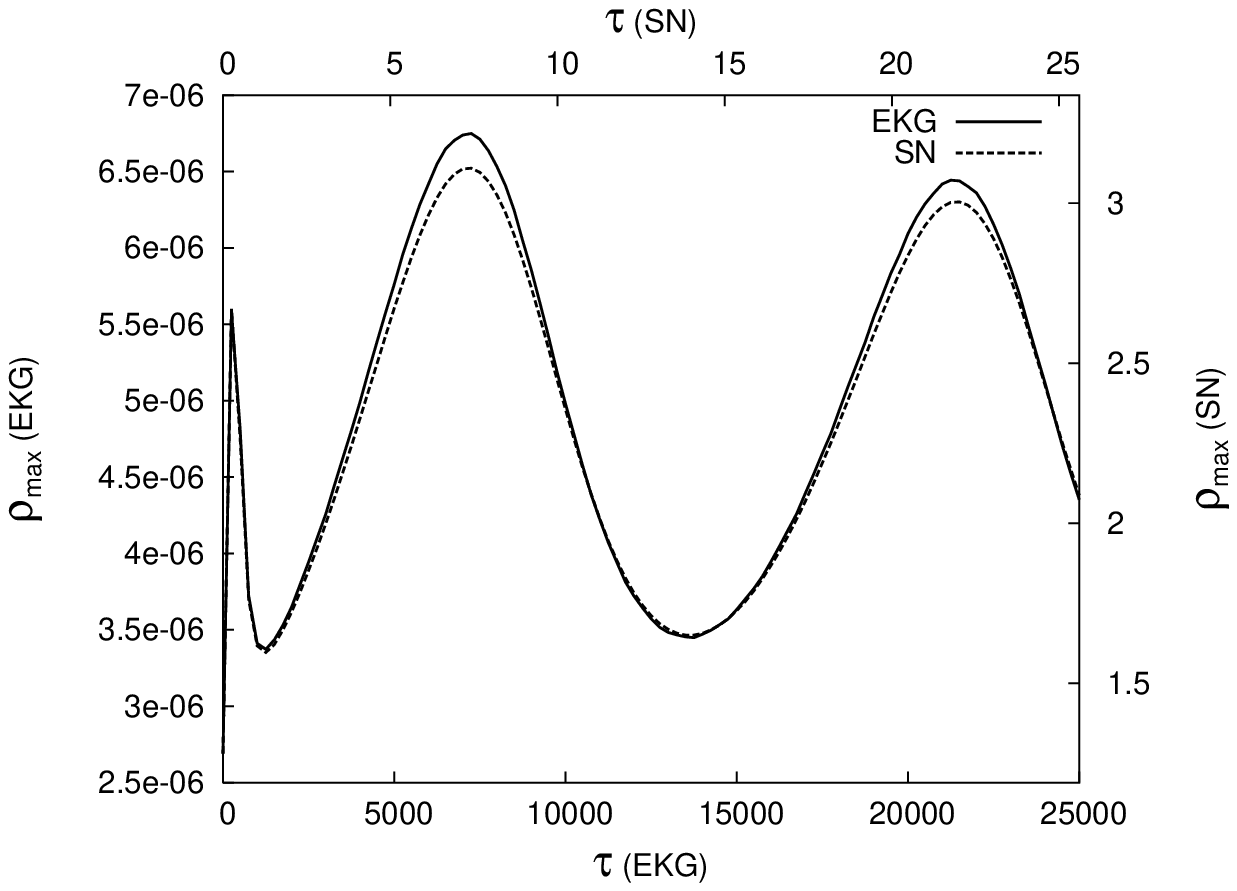}
\includegraphics[width=8cm]{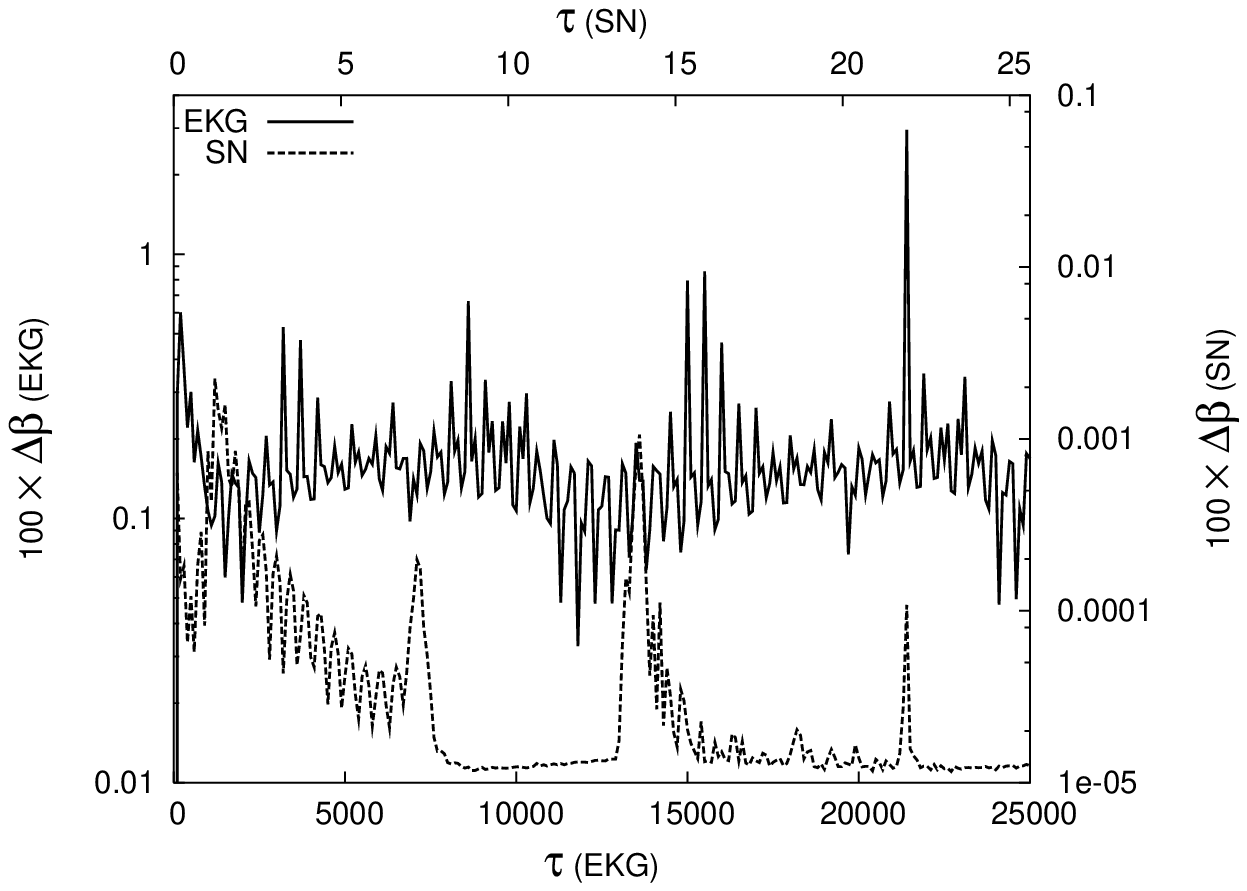}
\caption{\label{fig:scalinga} (Top) The numerical evolution of the maximum 
of the matter density $\rho_{max} (\tau)$ using the relativistic EKG system 
and a properly sized SN system. The initial scalar profile was a perturbed 
Newtonian oscillaton with $20$\% of mass excess, see text for details. (Bottom) Comparison of the relative violation of the momentum constraint $\Delta \beta$ for both the EKG\cite{phi2} and the SN system, see Eq.~(\ref{relbeta}). It can be seen that the Newtonian run is more accurate than the relativistic one; the reason for this difference is that the evolved system is too weak for the relativistic code. In both figures, the axes used for the relativistic system are labeled (EKG), while the label (SN) appears for the non--relativistic system. The \textit{output} data files for both systems are plotted without scaling, instead the scale transformation relating the runs can be calculated from the ranges shown in the axes. $\lambda=0.032$ for the case shown in here.}
\end{figure}

The resulting graphs of the energy density $\rho(\tau,0)$ for both the (post--Newtonian) SN and the (relativistic) EKG systems are shown in Fig.~\ref{fig:scalinga}. To begin with, a simple comparison of the scaled and non--scaled quantities tells us that the SN evolution is simpler and needs less computational effort when evolving weak field configurations. Also, we see from the plots that both evolutions are quite similar and then we can be confident that the SN system is also the weak--field approximation of oscillatons. 

The discrepancies seen in the graphs, which are less than $4$\%, can be attributed to the numerical error of the relativistic system (the system is too weak and the metric functions are pretty close to the values of the flat space, which makes them difficult to evolve accurately), and to the fact that we are at the edge of validity of the weak--field approximation. As it was first discussed in\cite{luis}, Newtonian oscillatons are valid if $\sqrt{8\pi G} |\Phi| = 2 \lambda ^2 \leq \sqrt{2} \times 10^{-3}$, and the scale parameter used in our example is just above this limit. Had we taken a weaker configuration, the EKG solution would have been even less accurate. 

In order to support the latter statement, we show in Fig.~\ref{fig:scalinga} the relative error $\Delta \beta$, Eq.~(\ref{relbeta}), for the relativistic\cite{phi2} and non-relativistic runs. Thus, we conclude that the SN result ($\Delta \beta \sim 10^{-7}$) is more accurate than the EKG one ($\Delta \beta \sim 10^{-3}$).

\begin{figure}[htp]
\includegraphics[width=8cm]{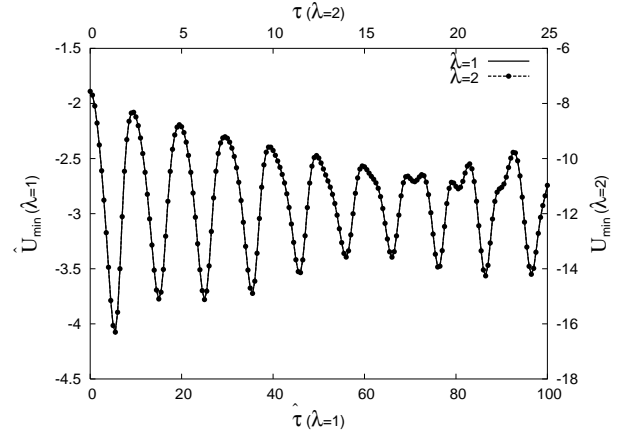}
\caption{\label{fig:scalingb} The minimum of the gravitational potential for the initial configuration $\hat{\psi}(\hat{x},\hat{0}) = e^{-(\hat{x}/2.75)^2}$ is shown. Superposed it is also the rescaled evolution for the corresponding configuration using $\lambda=2$, that is $\psi(x,0) = 4.0 e^{-(x/1.375)^2}$. The data files were not modified, and the ranges shown in the axis confirm the scaling properties of the SN system. For this case, $|U_\textrm{min}-\lambda^2 \hat{U}_\textrm{min}| < 2 \times 10^{-6}$, i.e., the difference is unnoticeable in the plot.}
\end{figure}

We have also evolved the SN system for different initial configurations related initially by a scaling transformation. An example of two Gaussians related initially by $\psi_0 = 4 \hat{\psi}_0$ is shown in Fig.~\ref{fig:scalingb}, where we plot the resulting gravitational potential. It can be seen that the scaling transformation is preserved always. In fact, for all cases we studied, the scaling transformation~(\ref{scale1}) was obeyed very accurately all along the evolution

\begin{figure}[htp]
\includegraphics[width=8cm]{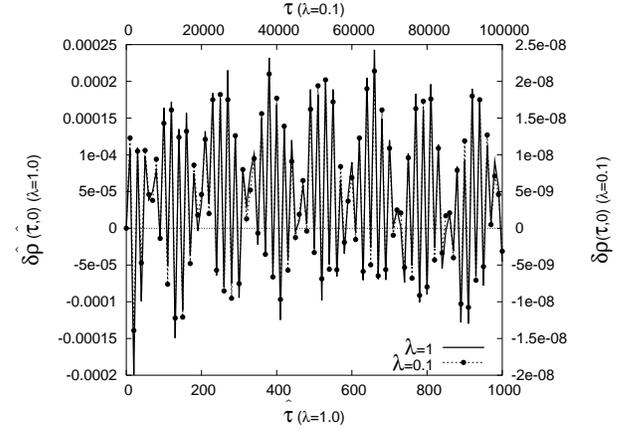}
\caption{\label{fig:scalingp} Scaling transformation of the density 
perturbations $\delta \rho (\tau,0)$ shown in Figs.~\ref{fig:test1} 
and \ref{fig:perts_ev}. The scale parameter used was $\lambda=0.1$. The 
data files were not modified, and the ranges shown in the 
axis confirm the scaling properties of the SN system. It is seen that 
the relation for the perturbation $\delta \rho = \lambda^4 \delta 
\hat{\rho}$ is satisfied, see Sec.~\ref{subsec:perturbations}, since
$|\delta \rho -\lambda^4 \delta \hat{\rho}| < 4 \times 10^{-9}$.}
\end{figure}

The scaling transformation of the perturbations in the energy density can also be seen in the evolved systems. In Fig.~\ref{fig:scalingp} we show again the perturbations of the configuration that appears in Figs.~\ref{fig:test1} and \ref{fig:perts_ev} (left axis), and the corresponding ones for the scaled system with $\lambda=0.1$ (right axis). As stated in Sec.~\ref{subsec:perturbations}, it is easily seen that the perturbations are related by the scaling transformation $\delta \rho = \lambda^4 \delta \hat{\rho}$.

\subsection{Equilibrium configurations in excited states}
Excited equilibrium configurations (also called $n$--node configurations) are also stationary solutions of the SN system, and the aim of this section is to investigate whether they are stable. What we find here is that, in general, all excited configurations are intrinsically unstable and decay onto 0--node configurations by emitting scalar matter. 

\begin{figure}[htp]
\includegraphics[width=8cm]{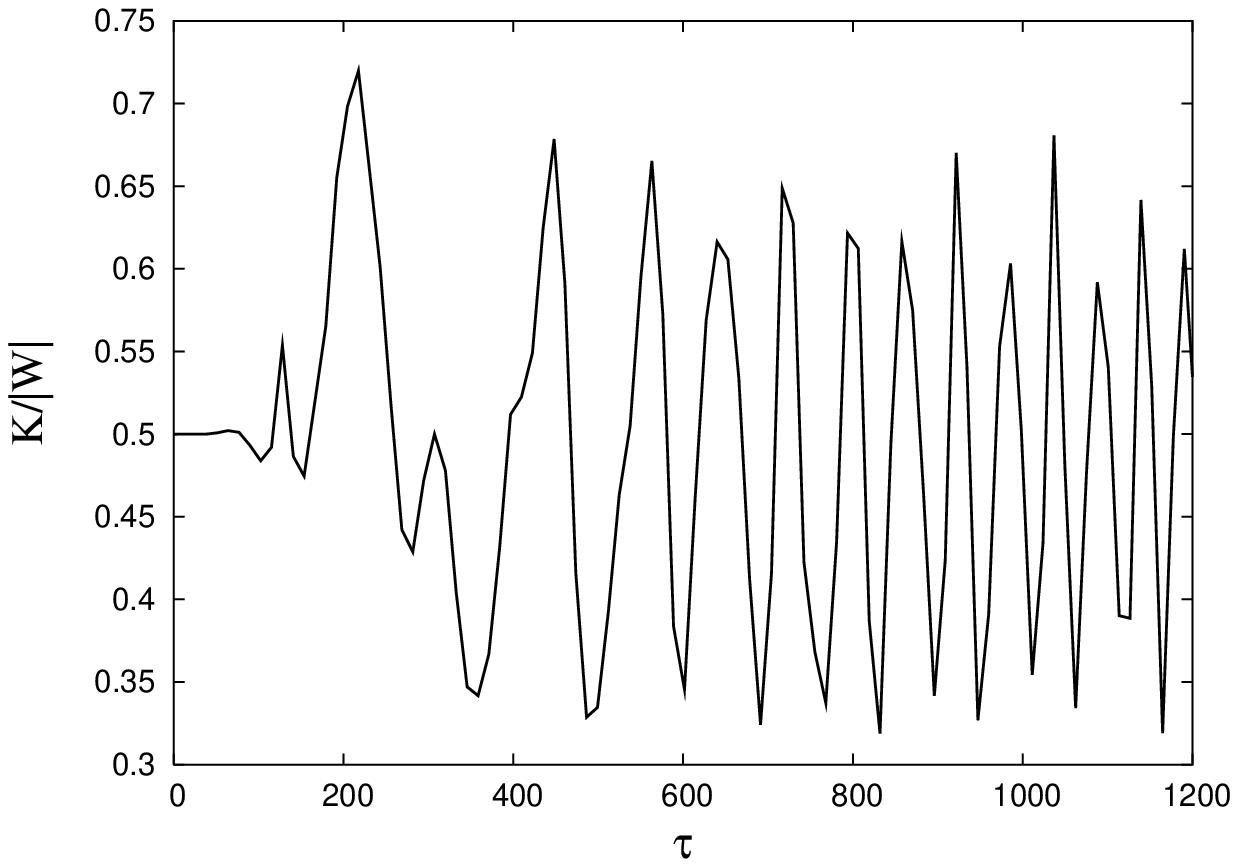}
\includegraphics[width=8cm]{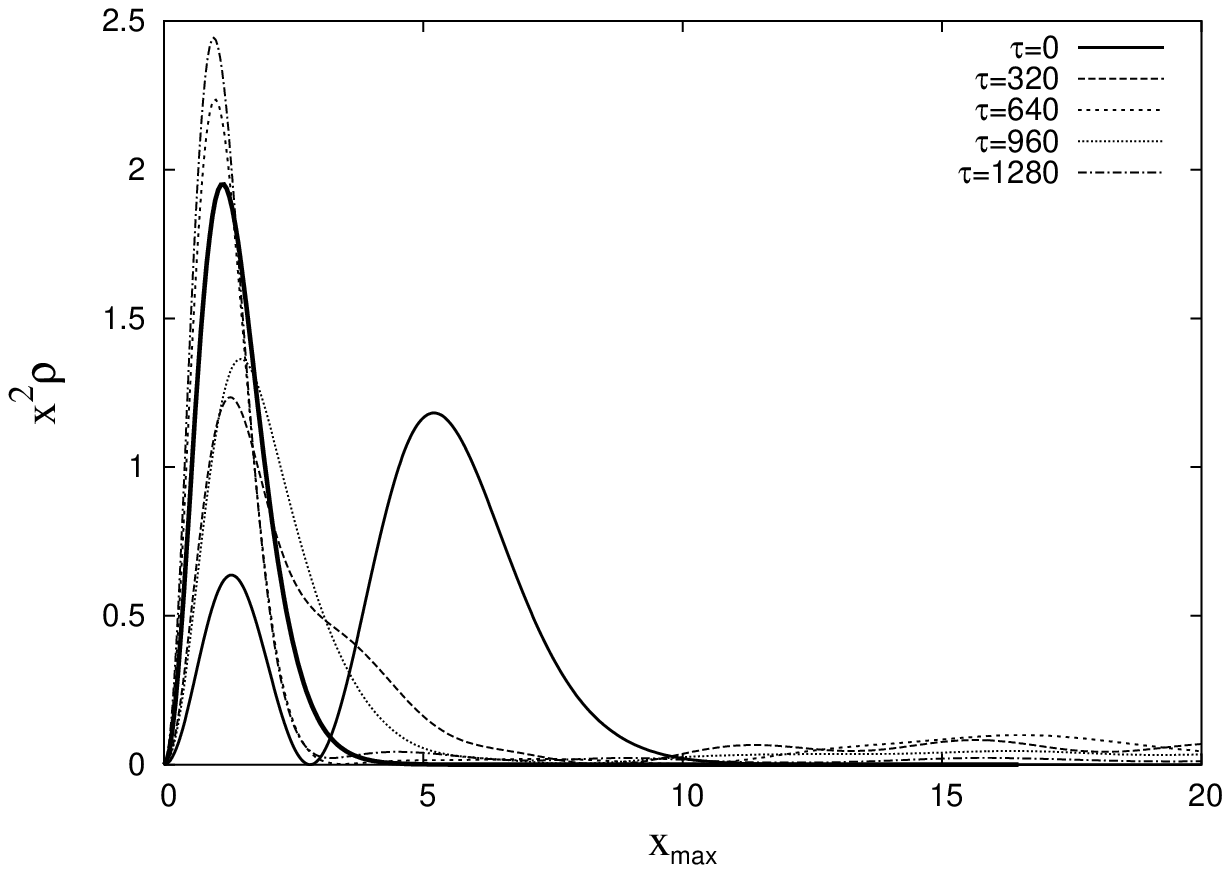}
\caption{\label{fig:excited} (Top) The virialization $K/|W|$ of a numerically evolved $1$--node equilibrium configuration. The system is initially virialized but it is nonetheless unstable. (Bottom) Some profiles of $x^2 \rho$ for a $1$--node equilibrium configuration at different times of its evolution. At the end of the run, the system has lost its node and evolves towards a $0$--node configuration represented by the thick solid curve, see also Fig.~\ref{fig:tracking}.}
\end{figure}

As a representative example of such decay, we show the numerical evolution of a $1$--node equilibrium configuration in Fig.~\ref{fig:excited}. Even though this excited configuration is initially virialized ($K/|W|=0.5$) and only perturbed by means of the discretization error, it ejects mass and settles down onto a 0--node equilibrium configuration, as it can be seen from the plots of the profiles of $x^2 \rho$ as the evolution proceeds. 

\begin{figure}[htp]
\includegraphics[width=8cm]{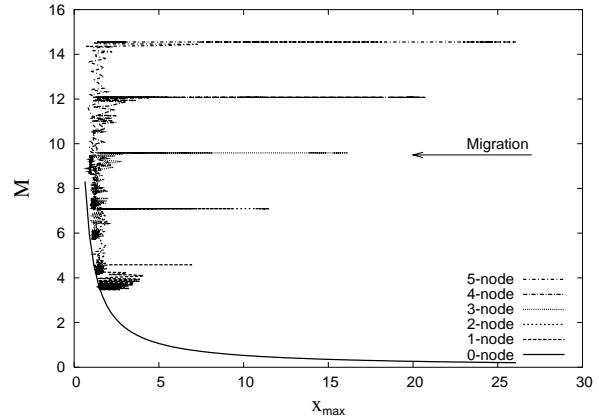}
\caption{\label{fig:tracking} 
Plots of the evolution of the mass number 
$M$ and $x_\textrm{max}$ for different (unperturbed) excited equilibrium configurations. The solid curve marks all possible $0$-node systems related through a scaling transformation, that is, those represented by the formula~(\ref{onodeformula}).}
\end{figure}

At later times, the energy density of the evolved $1$--node system oscillates with frequency $f=0.0976$, which means that it will settle down onto a $0$--node equilibrium configuration with a scale parameter $\lambda = \sqrt{f/\hat{f}}=1.457$, where $\hat{f}$ is the quasinormal mode found in Sec.~\ref{subsec:perturbations}. That is, the final state is that with $M=3.0$ and $x_\textrm{max}=1.77$. This last fact can be seen in a plot of $M$ vs $x_\textrm{max}$, as in Fig.~\ref{fig:tracking}, where the migration path of a $1$--node system can be followed until it ends up at a $0$--node equilibrium configuration.

The migration paths of different excited equilibrium configurations are also shown in Fig.~\ref{fig:tracking}. Without any added perturbation (apart from the discretization error), they all decay and migrate to a $0$--node solution. The latter are represented by the solid curve drawn by the formula 
\begin{equation}
M=M^\textrm{0--node} \left( \frac{x^\textrm{0--node}_\textrm{max}}{x_\textrm{max}} \right) \, , \label{onodeformula}
\end{equation}
with $M^\textrm{0--node}=2.0622$ and $x^\textrm{0--node}_\textrm{max}=2.58$. This last formula appears from the fact that the scale parameter is given by $\lambda=M/\hat{M}=\hat{x}_\textrm{max}/x_\textrm{max}$, see Eq.~(\ref{scale1}).

\section{Gravitational Cooling}
\label{sec:cooling}

In this section, we investigate an interesting issue which arises in the formation of gravitationally bounded objects, in particular, the scalar objects of this manuscript. 

It was discovered numerically\cite{seidel94} that, for the SN system, an arbitrary initial configuration settles down into a stable configuration through the emission of scalar matter, with the property that the ratio $K/|W|$ approaches and oscillates around 0.5 at late times, which means that the system is around virialization. 

This so--called gravitational cooling is so efficient, that allows the virialization of \textit{overwarmed} systems for which $K/|W| > 1$ initially. This is to be compared to the inability of the violent relaxation process to virialize such systems.

For the rest of this section, we study with more detail the gravitational cooling of arbitrary scalar systems whose evolution is dictated by the SN system. As we shall see, the results confirm the ability of the gravitational cooling to virialize \textit{all} possible initial configurations, even if they are too overwarmed. This section complements the relativistic studies in\cite{phi2}.

To begin with, we evolve the same configuration presented in\cite{seidel94}, but properly scaled in the form 
\begin{eqnarray}
\hat{\psi}(0,\hat{x}) = 4.5 e^{-(\hat{x}/2.1)^2}\left[ 1+0.5\sin(\pi \hat{x}/0.15) \right](1+i) \, ; \label{sands}
\end{eqnarray}
that is, we are using the scale parameter $\lambda=0.01$. The spatial and time resolutions were $\Delta \hat{x}=5\times 10^{-3}$ and $\Delta \hat{\tau} =1.25\times 10^{-5}$, respectively; and the run was followed up to $\hat{\tau}=6$ while the physical boundary was set at $\hat{x}_p=15$ (which correspond to $\tau=6\times 10^4$ and $x_p=1500$ as in\cite{seidel94}).

\begin{figure}[htp]
\includegraphics[width=8cm]{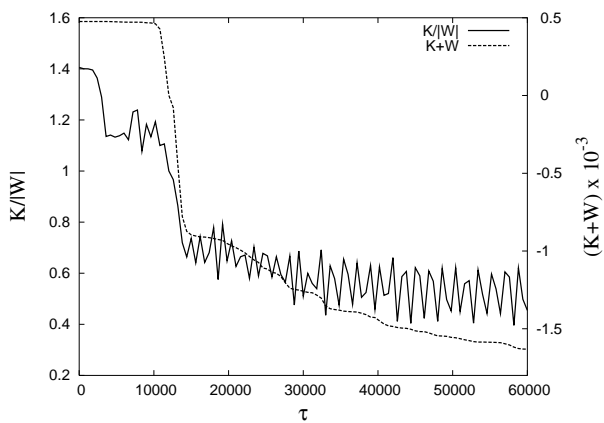}
\caption{\label{fig:cooling} The virialization $K/|W|$ (left axis) and total energy $E=K+W$ (right axis) for the initial distribution~(\ref{sands}). The graphs are scaled for an easy comparison with Fig.~3 in\cite{seidel94}.}
\end{figure}

\begin{figure}[htp]
\includegraphics[width=8cm]{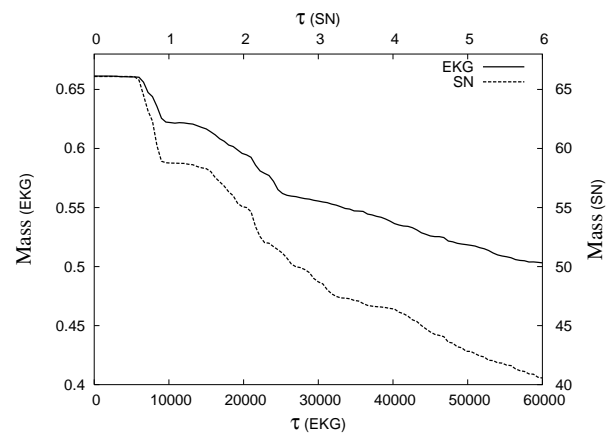}
\caption{\label{fig:sands} Comparison of the evolution of the total mass for 
the same Newtonian system shown in Fig.~\ref{fig:cooling} 
(see Eq.~(\ref{sands})), and the relativistic system represented by 
Eqs.~(\ref{psi_rel}), see text for details. As in 
Figs.~\ref{fig:scalinga} and~\ref{fig:scalingb}, the output data files for both systems are plotted without modification, and the scale transformation relating the runs can be calculated from the ranges shown in the axes. $\lambda=0.032$ for the case shown in here.}

\end{figure}

In Fig.~\ref{fig:cooling}, we show the same quantities shown in Fig.~3 of\cite{seidel94}, which are the virialization $K/|W|$ and the total energy of the system $E=K+W$; the corresponding graph of the mass is shown in Fig.~\ref{fig:sands} and will be discussed later. Even though the results agree qualitatively, we find that the values of the mass and the total energy at the end of the run are not the same.

The system seems to settle down onto an equilibrium configuration with a (non-scaled) mass of around $M=0.45$. But, this value is within the relativistic realm, since for Newtonian equilibrium configurations $M < 0.08$. Thus, we consider it would be more appropriate to evolve the system~(\ref{sands}) using the relativistic EKG equations.

We made a numerical run of the EKG system for a real scalar field, using the following initial profiles for the scalar field and its time derivative, respectively,
\begin{subequations}
\label{psi_rel}
\begin{eqnarray}
\sqrt{\kappa_0}\Phi^\textrm{(r)}(0,x)&=&2\lambda^2 \textrm{Re}(\hat{\psi}(0,x)) \, , \\
\sqrt{\kappa_0}\frac{\partial {\Phi}^\textrm{(r)}}{\partial \tau}(0,x)&=&2\lambda^2 \textrm{Im}(\hat{\psi}(0,x)) \, ,
\end{eqnarray}
\end{subequations}
as follows from Eq.~(\ref{phi_expansion}) and the post--Newtonian rules. The spatial and time resolutions of this relativistic evolution were $\Delta x=0.5$ and $\Delta \tau=0.01$, respectively, while the numerical boundary was set at $x_N=1500$.

For comparison, we plot in Fig.~\ref{fig:sands} the total mass of the SN system with that of the EKG system. Initially, both systems have the same mass, but the path followed by each system is different as the evolution proceeds. As we mentioned before, we believe that the true final state is that of the relativistic system, which in this cases corresponds to a relativistic oscillaton. 

At this point, we would like to mention that the comparison between violent relaxation and the gravitational cooling is not appropriate in this example. As Fig.~\ref{fig:sands} shows, the initial configuration is so massive that it cannot disperse away because of the intervention of strong gravitational effects fully accounted by the Einstein equations, despite the large initial ratio $K/|W|=1.4$. Actually, the inclusion of relativistic effects can prevent the appearance of the gravitational cooling and make the systems collapse into a black hole\cite{phi2}.

Nevertheless, we have indeed found that the gravitational cooling is a very efficient mechanism for Newtonian systems (for the relativistic case see\cite{seidel94,phi2}) even in the case in which $K/|W| > 1$. 

\begin{figure}[htp]
\includegraphics[width=8cm]{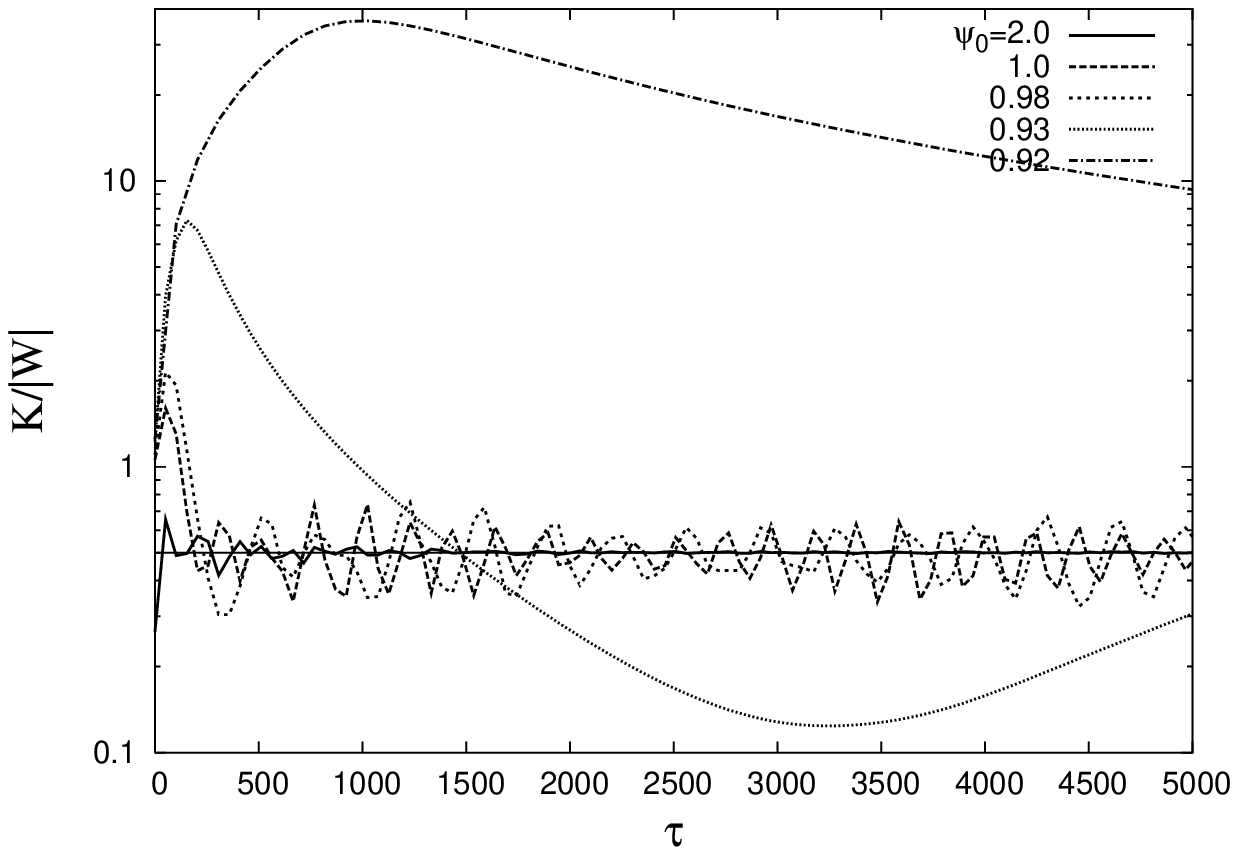}
\includegraphics[width=8cm]{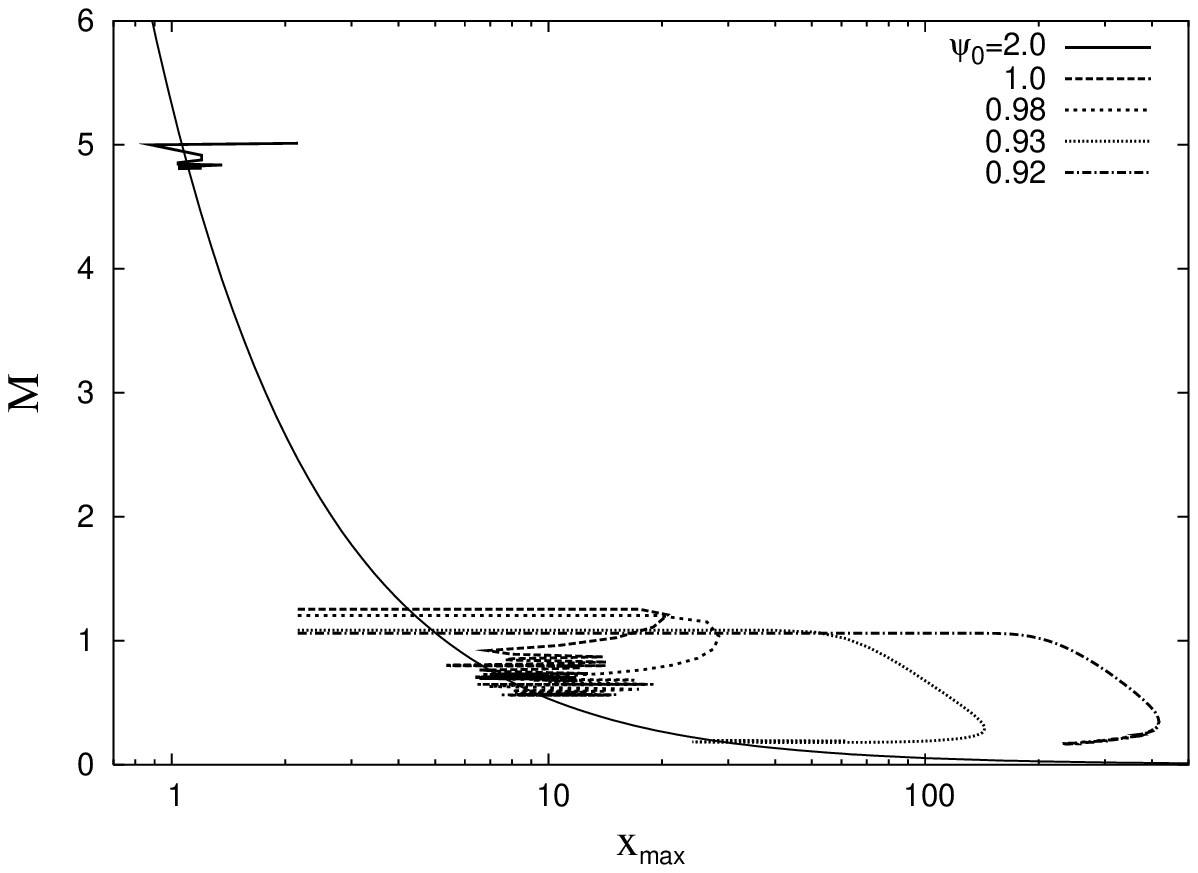}
\caption{\label{fig:coolnewt} 
Evolution of different overwarmed and 
overcooled Gaussian profiles of the form $\psi(0,x)=\psi_0 e^{-(x/2)^2}$. Shown are the cases $\psi_0=2.0,1.0,0.98,0.93,0.92$. (Top) Independently of the initial value of $K/|W|$, the scalar systems always collapse back to an equilibrium configuration. The virialization of the systems commences once $K/|W| \sim 0.5$. (Bottom) Migration paths followed by the aforementioned systems in a $M$ vs $x_\textrm{max}$ plot. The paths converge to the line representing $0$--node equilibrium configurations, see Fig.~\ref{fig:tracking} and Eq.~(\ref{onodeformula}).}
\end{figure}

To investigate this, we have made runs with an initial Gaussian profile of the form $\psi(0,x)=\psi_0 e^{-(x/2)^2}$ taking different values values of $\psi_0$. In Fig.~\ref{fig:coolnewt} we show the graphs of the virialization coefficient $K/|W|$ and the corresponding path in a $M$ vs $x_\textrm{max}$ plot of the evolved Gaussian profiles.

For the value $\psi_0=2.0$ ($K/|W|_{\tau=0}=0.265$), the system rapidly virializes, and, from $\tau=1500$ afterwards, it has settled down onto a $0$--node equilibrium configuration corresponding to a scale parameter $\lambda=2.33$. For the values $\psi_0=1.0$ ($K/|W|_{\tau=0}=1.06$) and $\psi_0=0.98$ ($K/|W|_{\tau=0}=1.104$), it is noticed that the systems are approaching to $0$--node equilibrium configurations, but it will take a longer time for them to virialize completely.

The examples with $\psi_0=0.93$ ($K/|W|_{\tau=0}=1.225$) and $\psi_0=0.92$ ($K/|W|_{\tau=0}=1.252$) illustrate how long it takes for a system to virialize as we consider lower values of $\psi_0$. For instance, for $\psi_0=0.92$, the system will need much more time than shown in Fig.~\ref{fig:coolnewt} to begin to virialize, but we can anticipate that it will follow a similar path as $\psi_0=0.93$. Actually, the migration paths of these systems were obtained by evolving them up to times of the order $\tau > 10^4$.

These examples show that all \textit{overcooled} profiles ($K/|W| < 0.5$) 
rapidly settle down onto a $0$--node equilibrium configuration, but 
overwarmed profiles require much longer times to 
stabilize, a fact that makes them difficult to evolve numerically.

Despite of this, we conclude that if the overwarmed runs are followed during a sufficient long time, they will always find their way to an appropriate $0$--node equilibrium configuration. The reason for this is that there is an infinite number of $0$--node equilibrium configurations, including those in which the scale parameter is too small, $\lambda \ll 1$; that is, there is always an equilibrium configuration available for any evolved initial profile.

\section{Conclusions}
\label{conclusions}
In this paper, we have studied the evolution of a self-gravitating scalar field in the Newtonian regime using numerical methods. The latter were systematically tested for their accuracy, convergence and reliability. In all cases, the numerical code gave the expected correct results of the $0$--node equilibrium configurations and its perturbations. Also, the numerical code preserved the scaling invariance of the SN system all along the time of the evolutions.

An important point was the boundary condition imposed on the SN system. We found that the implementation of a sponge by adding an imaginary potential is an appropriate boundary condition. It allowed us to maintain under control the amount of scalar matter reflected by the numerical boundary.

Our results imply that $0$--node equilibrium configurations are intrinsically stable against radial perturbations, and that they play the important role of final states arbitrary scalar systems settle down onto in the Newtonian regime. On the contrary, excited equilibrium configurations were found to be intrinsically unstable configurations. Even though they are initially virialized, they evolve towards a $0$-node solution.

An important point here is that, due to the scaling properties of the SN system, the study of the whole space of possible equilibrium configurations was reduced to the analysis of some properly sized configurations. Moreover, the same was done to study non-equilibrium configurations to give simplified and accurate simulations.

Last but not least, we retook the analysis of the gravitational cooling within the Newtonian regime of scalar fields. The main result is that, in the weak field limit, the gravitational cooling is a very efficient mechanism, which allows any initial configurations to decay into a $0$--node Newtonian scalar soliton. So far, we have not found any evidence for systems that disperse away by ejecting all their mass.

We expect that the results presented here would provide useful information about structure formation in the universe, not only regarding models of scalar field dark matter as in\cite{fsglau,sin,hu,arbey,gal}, but also about other models whose dynamics is governed by the SN system beyond spherical symmetry as in\cite{schmethod}, case about which we expect to report in the near future.


\begin{acknowledgments}
We are thankful to Miguel Alcubierre, Horst Beyer, Andrew Liddle, Ed Seidel and Jonathan Thornburg for fruitful discussions. L.A.U-L. acknowledges the kind hospitality of the AEI where part of this work was developed, and the support from PROMEP and Guanajuato University projects. This research is partly supported by the bilateral project DFG-Conacyt 444-113/16/3-1.
\end{acknowledgments}



\begin{thebibliography}{}
\bibitem{fsglau} F. S. Guzm\'an and L. A. Ure\~na-L\'opez, Phys. Rev. D {\bf 68}        , 024023 (2003), astro-ph/0303440.
\bibitem{schmethod} G. Davies and L. M. Widrow, Astrophys. J. {\bf 485}, 484            (1997), astro-ph/9607133.
\bibitem{seidel90} E. Seidel and W-M. Suen, Phys. Rev. D {\bf 42}, 384 (1990).          J. Balakrishna, E. Seidel, and W-M. Suen, Phys. Rev. D {\bf 58}, 104004         (1998), gr-qc/9712064.
\bibitem{seidel94} E. Seidel and W-M. Suen, Phys. Rev. Lett. {\bf 72}, 2516             (1994), gr-qc/9309015.
\bibitem{hu} W. Hu, R. Barkana and A. Gruzinov, Phys. Rev. Lett. \textbf{85},           1158 (2000), astro-ph/0003365.
\bibitem{phi2} M. Alcubierre, R. Becerril, F. S. Guzm\'an, T. Matos, D.                 N\'u\~nez and L. A. Ure\~na-L\'opez, Class. Quantum Grav. {\bf 20},             2883 (2003), gr-qc/0301105.
\bibitem{dmota} D. Mota and  C. van de Bruck, \textit{On the spherical collapse model in dark energy cosmologies}, astro-ph/0401504.
\bibitem{jcerv}  M. A. Rodriguez-Meza and J. L. Cervantes-Cota, \textit{Potential--density pairs for spherical galaxies and bulges: the influence of scalar fields}, astro-ph/0401572.
\bibitem{fuchs}  B. Fuchs and E. W. Mielke, \textit{Scaling behaviour of a scalar field model of dark matter halos}, astro-ph/0401575.
\bibitem{hawley1} S. H. Hawley and M. W. Choptuik, Phys. Rev. D {\bf 62},               104024 (2000), gr-qc/0007039. 
\bibitem{hawley} S. H. Hawley and M. W. Choptuik, Phys. Rev. D {\bf 67}, 024010         (2003), gr-qc/0208078.
\bibitem{ruffini} R. Ruffini and S. Bonazzola, Phys. Rev. {\bf 187}, 1767
	(1969).
\bibitem{colpi} M. Colpi, S. L. Shapiro and I. Wasserman, Phys. Rev. Lett.              {\bf 57}, 2485 (1986).
\bibitem{liddle} A. R. Liddle and M. S. Madsen, Int. J. Mod. Phys. {\bf D 1},           101 (1992).
\bibitem{pang} R. Friedberg, T. D. Lee, and Y. Pang, Phys. Rev. D {\bf 35}, 
	3640 (1987).
\bibitem{lee} J. W. Lee and I. G. Koh, Phys. Rev. D {\bf 53}, 2236 (1996), 
	hep-ph/9507385.
\bibitem{moroz} I. M. Moroz, R. Penrose, and P. Tod, Class. Quantum Grav.
	{\bf 15}, 2733 (1998); P. Tod and I. M. Moroz, Nonlinearity {\bf 12},
	201 (1999).
\bibitem{gleiser} M. Gleiser, Phys. Rev. D {\bf 38}, 2376 (1988).

\bibitem{mielke} E. W. Mielke and F. E. Schunck, {\it Proceedings of the 8th
	Marcel Grossmann Meeting}, Jerusalem, Israel (World Scientific,
	Singapore, 1999), gr-qc/9801063.
\bibitem{diegos} F. E. Schunck and D. F. Torres, Int. J. Mod. Phys. {\bf D 9},        601 (2000).
\bibitem{silveira} V. Silveira and C. M. G. de Sousa, Phys. Rev. D {\bf 52},            5724 (1995), astro-ph/9508034.
\bibitem{yoshida} S. Yoshida and Y. Eriguchi, Phys. Rev. D {\bf 55}, 1994               (1997). S. Yoshida and Y. Eriguchi, Phys. Rev. D {\bf 56}, 762 (1997).
\bibitem{ryan} F. D. Ryan, Phys. Rev. D {\bf 55}, 6081 (1997).
\bibitem{kobayashi} Y-S Kobayashi, M. Kasai and T. Futamase, Phys. Rev. D               {\bf 50} 7721 (1994).
\bibitem{jetzer} P. Jetzer, Phys. Rept. {\bf 220} 163 (1992).
\bibitem{mielkes} F. E. Schunck and. E. W. Mielke, Class. Quantum Grav.                 {\bf 20}, R301 (2003).
\bibitem{seidel91} E. Seidel and W-M. Suen, Phys. Rev. Lett. {\bf 66}, 1659
	(1991).
\bibitem{luis} L. A. Ure\~na-L\'opez, Class. Quantum Grav. {\bf 19}, 2617
	(2002), gr-qc/0104093; L. A. Ure\~na-L\'opez, T. Matos, and 
	R. Becerril, Class. Quantum Grav. \textbf{19}, 6259 (2002).
\bibitem{dpage} D. Page, \textit{Classical and Quantum decay of 
oscillatons: oscillating self-gravitational real scalar field solitons}, 
gr-qc/0310006.

\bibitem{nr} W. H. Press, S. A. Teukolsky, W. T. Watterling and
  B. P. Flanery,          {\it Numerical Recipies in fortran}
  (Cambridge University Press, 1996).
\bibitem{cphysics} S. E Koonin and D. C. Meredith, {\it Computational Physics}
        (Addison-Wesley Publishing Company, 1990). 
\bibitem{israeli} M. Israeli, and S.A. Orszag, J. Comp. Phys. \textbf{41}, 115          (1981).
\bibitem{sin} S. J. Sin, Phys. Rev. D {\bf 50}, 3650 (1994), {\tt
	hep-ph/9205208}; S. U. Ji and S. J. Sin, Phys. Rev. D {\bf 50}, 3655            (1994), hep-ph/9409267.
\bibitem{gal} M. Alcubierre, F. S. Guzm\'an, T. Matos, D. N\'u\~nez, L. A.              Ure\~na-L\'opez and P. Wiederhold, Class. Quantum Grav. {\bf 19}, 5017          (2002), gr-qc/0110102.
\bibitem{arbey} A. Arbey, J. Lesgourgues, and P. Salati, Phys. Rev. D {\bf 64},         123528 (2001), astro-ph/0105564; {\it ibid}, {\bf 65}, 083514 (2002), astro-ph/0112324; \textit {ibid}, \textbf{68}, 023511 (2003), astro-ph/0301533.

\end{thebibliography}
\end{document}